\documentclass[aps,twocolumn,amsmath,amssymb,showpacs,prb,eqsecnum,10pt]{revtex4-1}
\usepackage{amsmath,color,colordvi}
\usepackage[pdftex]{graphicx}
\usepackage{dcolumn}% Align table columns on decimal point
\usepackage{bm}% bold math
\usepackage{dsfont}
\usepackage{color}

\usepackage[english]{babel}

\begin{document}

\title{Effect of the electromagnetic environment on current fluctuations\\ in driven tunnel junctions}

\author{Moritz Frey and Hermann Grabert}

\affiliation{Physikalisches Institut, Universit\"at
Freiburg, Hermann-Herder-Stra{\ss}e 3, 79104 Freiburg, Germany}

\begin{abstract}
We examine current fluctuations in tunnel junctions driven by a superposition of a constant and a sinusoidal voltage source. In standard setups the external voltage is applied to the tunneling element via an impedance providing an electromagnetic environment of the junction. The modes of this environment are excited by the time-dependent voltage and are the source of Johnson-Nyquist noise. We determine the autocorrelation function of the current flowing in the leads of the junction in the weak tunneling limit up to terms of second order in the tunneling Hamiltonian. The driven modes of the electromagnetic environment are treated exactly by means of a unitary transformation introduced recently. Particular emphasis is placed on the spectral function of the current fluctuations.
The spectrum is found to comprise three contributions: a term arising from the Johnson-Nyquist noise of the environmental impedance, a part due to the shot noise of the tunneling element and a third contribution which comes from the cross-correlation between fluctuations caused by the electromagnetic environment and fluctuations of the tunneling current. All three parts of the spectral function occur already for devices under dc bias. The spectral function of ac driven tunneling elements can be determined from the result for a dc bias by means of a photo-assisted tunneling relation of the Tien-Gordon type. Specific results are given for an Ohmic environment and for a junction driven through a resonator.
\end{abstract}

\date{29 February 2016, Published in: Phys.\ Rev.\ B {\bf 94}, 045429 (2016)}


\maketitle

\section{Introduction}\label{sec:one}
The effect of the electromagnetic environment on tunnel junctions has extensively been studied some 25 years ago both experimentally\cite{Delsing_1989,Geerligs_1989,Cleland_1992,Holst_1994} and theoretically.\cite{Devoret_1990,Girvin_1990,Grabert_1991,Ingold_1992,Lee_1996} The theory of the dynamical Coulomb blockade (DCB), frequently also referred to as $P(E)$ theory, has explained the experimentally observed suppression of the tunneling current at low voltage bias as an effect of photon exchange between the tunneling element and its electromagnetic environment. 
More recently, new experiments\cite{Hofheinz_2011,Altimiras_2014,Parlavecchio_2015} with designed, approximately single-mode electromagnetic environments have led to a revival of the DCB theory.\cite{Safi_2011,Souquet_2013,Grabert_2015,Roussel_2016,Mora_2016} In particular, in view of advances in microwave technology, ac driven tunneling elements have moved to the focus of attention.

The study of ac driven devices is extensive and has been reviewed by several authors. \cite{Tucker_1985,Platerno_2004,Brandes_2005} Frequently, also properties of the current noise \cite{Blanter_2000} have been addressed in this context. In this paper the focus is on the effects of the electromagnetic environment on the current noise of an ac driven tunnel junction. Previous work  mostly assumes that the external driving leads to a time dependence of properties of the tunneling system itself\cite{Platerno_2004,Brandes_2005} rather than studying the driving by an external voltage source connected to the tunneling system via leads of finite impedance. The work by Safi and coworkers\cite{Safi_2010,Safi_2014} presents a rather general approach to ac driven systems and includes an environmental impedance. However, the analysis is then based on the assumption that in the absence of tunneling the charge of the tunneling system is conserved. This means that displacement currents are not taken into account. These currents are, however, crucial to describe the influence of the electromagnetic environment accurately. 
 
The standard Hamiltonian model of conventional DCB theory properly describes a tunneling element biased by a voltage source via an environmental impedance. In the original work\cite{Devoret_1990,Girvin_1990,Grabert_1991,Ingold_1992} this model was only studied for applied dc voltages. We have recently shown\cite{Grabert_2015} that the analysis of the model for ac bias voltages entails substantial modifications of the theory. The time dependence of the Hamiltonian arising from the external bias can no longer be transformed in the usual way\cite{Ingold_1992} into a time-dependent phase factor of the tunneling Hamiltonian. This is due to the fact that the modes of the electromagnetic environment are excited by an alternating voltage source and constitute a driven quantum bath.\cite{Grabert_2016} 

A suitable way of handling time-dependent voltages within $P(E)$ theory is based on a unitary transformation involving also the environmental degrees of freedom.\cite{Grabert_2015} Employing this method, the average current flowing through the environmental impedance into the outer circuit was determined, and for ac driven devices a suppression of higher harmonics of the current by the electromagnetic environment was found.\cite{Grabert_2015} In this paper we now investigate specifically the autocorrelation function of the current flowing in the leads of a tunneling element driven by dc and ac voltage sources. The experimentally relevant spectral density of current fluctuations is determined and discussed for specific models of the electrodynamic environment.

The paper is organized as follows. In Sec.~\ref{sec:two} we recall the standard Hamiltonian of a voltage biased tunnel junction with an electrodynamic environment characterized by a lead impedance. The relevant current operators are introduced and the perturbation theory in the tunneling Hamiltonian is outlined. Also, the unitary transformation removing the time dependence of the environmental Hamiltonian is specified. Section \ref{sec:three} introduces the current autocorrelation function. All terms up to second order in the tunneling Hamiltonian contributing to the current correlator are determined. Each of these terms factorizes into an average over the quasiparticles in the leads of the tunnel junction and an average over the electromagnetic environment. It is shown that for arbitrary driving voltages these averages can be fully expressed in terms of quantities known from the standard $P(E)$ theory for constant voltage bias. Finally, the results are specified for a driving voltage comprising a dc voltage and a sinusoidal ac voltage.

In Sec.~\ref{sec:four} we switch to Fourier space and introduce the Fourier coefficients of the steady state current autocorrelation function of a periodically driven junction.  These Fourier coefficients are functions of the lag time, that is the time difference between the positions in time of the two current operators linked by the correlation function. The zero order Fourier coefficient describes the time-averaged current correlator where the absolute positions in time of the two current operators are averaged at constant lag time over one period of the driving voltage. We then introduce the Fourier transform of the time-averaged current correlator which gives the experimentally relevant spectral function. 

The explicit evaluation of the spectral function naturally divides into three steps. First, we determine the part of the spectrum coming from the Johnson-Nyquist noise of the environmental impedance. Next, we calculate the contribution to the spectrum arising from the shot noise of the tunneling current. In addition, we obtain a third part of the spectrum which is due to the cross-correlation between fluctuations of the tunneling current and Johnson-Nyquist voltage fluctuations across the environmental impedance.  

In Sec.~\ref{sec:five} we illustrate our findings by studying specific models of the electromagnetic environment. First, we consider the case of a strictly Ohmic environment with a constant environmental impedance. In the limit of a low impedance environment one recovers the shot noise of a tunnel junction in the absence of DCB effects. As a second example we study a tunnel junction driven through an $LC$-resonator. At the resonance frequency the noise is strongly enhanced and the cross-correlation part of the spectrum predicted in Sec.~\ref{sec:four} is shown to be very significant. Finally, in Sec.~\ref{sec:six} we present our conclusions.

%%%%%%%%%%%%%%%%%%%%%%%%
%%%%%%%%%%%%%%%%%%%%%%%%
%%%%%%%%%%%%%%%%%%%%%%%%
\section{Model and perturbation theory}\label{sec:two}

\begin{figure}
\includegraphics[width=0.4\textwidth]{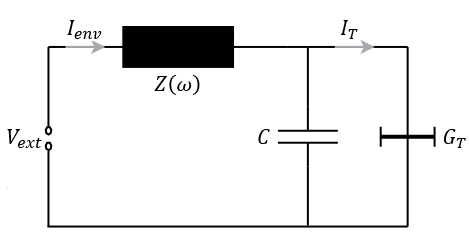}
\caption{\label{fig1} Circuit diagram of a voltage biased tunneling element showing a tunnel junction with capacitance $C$ and tunneling conductance $G_T$ coupled to a voltage source $V_{ext}$ via an external impedance $Z(\omega)$. The current $I_{env}$ flowing through the environmental impedance and the tunneling current $I_T$ are also indicated.}
\end{figure}

We consider the standard model for the DCB\cite{Devoret_1990,Girvin_1990,Grabert_1991,Ingold_1992}, a tunnel junction with junction capacitance $C$ and tunneling conductance $G_T$  driven by a voltage source $V_{ext}(t)$ via an environmental impedance $Z(\omega)$. A circuit diagram of the setup indicating also the currents flowing in the circuit is shown in Fig.~\ref{fig1}.

%%%%%%%%%%%%%%%%%%%%%%%
\subsection{Model Hamiltonian}
The Hamiltonian of a voltage biased tunnel junction may be written as
\begin{equation}\label{Htot}
H=H_{el} +H_T + H_{env}
\end{equation}
where $H_{el}$ describes the conduction electrons of the leads on either side of the tunnel junction
\begin{equation} \label{Hqp}
H_{el}=\sum_{k,\sigma}\epsilon_{k\sigma}a_{k\sigma}^{\dag}a_{k\sigma}^{}+
      \sum_{q,\sigma}\epsilon_{q\sigma}a_{q\sigma}^{\dag}a_{q\sigma}^{}\, .
\end{equation}
The operator $a_{k\sigma}^{}$ [$a_{q\sigma}^{}$] is the annihilation operator of an electron state with energy $\epsilon_{k\sigma}$ [$\epsilon_{q\sigma}$] in the left [right] electrode, where $k$ [$q$] denotes the longitudinal wave number and $\sigma$ denotes the transversal and spin quantum numbers.  The quantum number $\sigma$ is conserved during tunneling transitions described by the tunneling Hamiltonian
\begin{equation}\label{HT}
 H_T=\Theta \, e^{-i\varphi} + \Theta^{\dagger}e^{i\varphi}
\end{equation}
with the quasiparticle tunnel operator
\begin{equation}\label{tunop}
\Theta =   \sum_{k,q,\sigma} t_{kq\sigma} a_{k\sigma}^{\dag}a_{q\sigma}^{}
\end{equation}
where $t_{kq \sigma}$ is the tunneling amplitude. The phase operator $\varphi$  is conjugate to the junction charge $Q$ and obeys the canonical commutation relation\cite{Devoret_1990}
$
\left[ \varphi, Q\right] = ie 
$.
The last term in Eq.~(\ref{Htot}) describes the junction capacitance $C$ and the environmental impedance $Z(\omega)$ and it may be written in the form\cite{Ingold_1992}
\begin{eqnarray}\label{Henv}
&&H_{env}(t) =\frac{Q^2}{2C}\\ \nonumber
&&\qquad +\sum_n \left\{\frac{Q_n^2}{2C_n}  +\frac{1}{2L_n} \left(\frac{\hbar}{e}\right)^2\left[\varphi-\varphi_n-\varphi_{ext}(t)\right]^2  \right\}
\end{eqnarray}
where the first term is the charging energy of the tunnel junction with capacitance $C$ and
the second term represents the impedance $Z(\omega)$  as a collection of $LC$ circuits.\cite{Caldeira_1981,Caldeira_1983} These environmental modes form a thermal bath causing dissipation and fluctuations and lead to the DCB effect. The charge operators $Q_n$ are conjugate to the phase operators $\varphi_n$ with the commutators
$
\left[\varphi_n, Q_n \right] = ie 
$, and the phase 
\begin{equation}\label{phiext}
\dot \varphi_{ext}(t) = \frac{e}{\hbar} V_{ext}(t)
\end{equation}
describes the external driving where $V_{ext}(t)$ is the voltage applied to the circuit.

%%%%%%%%%%%%%%%%%%%%%%%
\subsection{Current operators}
Using the Hamiltonian (\ref{Htot}), we find for the time rate of change of the junction charge
\begin{equation}
\dot Q = \frac{i}{\hbar}\left[ H,Q\right] = I_{env}-I_T
\end{equation}
where\cite{Lee_1996,Goppert_2000}
\begin{equation}\label{Ienv}
I_{env}=\frac{i}{\hbar}\left[ H_{env},Q\right] =-\frac{\hbar}{e}\sum_n \frac{\varphi-\varphi_n-\varphi_{ext}}{L_n} 
\end{equation}
is the current flowing through the environmental impedance $Z(\omega)$  while
\begin{equation}\label{Itun2}
I_T=- \frac{i}{\hbar}\left[ H_T,Q\right]=- i\frac{e}{\hbar}\left[\Theta\, e^{-i\varphi}- \Theta^{\dagger}\, e^{i\varphi}\right]
\end{equation}
is the tunneling current flowing across the tunneling element. Tunneling currents are not experimentally accessible\cite{Landauer_1991} since measurements typically relate to the  current $I_{env}$ entering the outer circuit.

%%%%%%%%%%%%%%%%%%%%%%%%
\subsection{Perturbation theory in the tunneling Hamiltonian}
To evaluate the time evolution for weak tunneling we write the total Hamiltonian (\ref{Htot}) as
$
H=H_0+H_T
$
where
$
H_0=H_{el}+H_{env}
$
is the Hamiltonian in the absence of tunneling. A straightforward expansion in terms of $H_T$ yields
for the Heisenberg current operator $I_{env}(t)$ up to terms of second order in $H_T$
\begin{eqnarray}\label{IenvHeis}\nonumber
&&I_{env}(t) = \check I_{env}(t)   \\ \nonumber
&&\qquad
+\frac{i}{\hbar} \int_{0}^t ds\,   \left[\check\Theta(s)\, e^{-i\check\varphi(s)}+ \hbox{H.c.}\, ,  \check I_{env}  (t)\right] \\ \nonumber
&&\qquad -\frac{1}{\hbar^2}\int_{0}^t ds_1 \int_{0}^{s_1} ds_2\,   \Big[\check\Theta(s_2)\, e^{-i\check\varphi(s_2)}+ \hbox{H.c.}\, , \\ 
&&\qquad\qquad \Big[\check\Theta(s_1)\, e^{-i\check\varphi(s_1)}+ \hbox{H.c.}\, ,  \check I_{env}  (t)\Big] \Big].
\end{eqnarray}
Here we have introduced the interaction representation of operators
$
\check A(t)=U_0^{\dagger}(t,0)\, A\, U_0(t,0)
$ with the unperturbed time evolution operator
\begin{equation}\label{U0}
U_0(t,t^{\prime}) = \mathcal{T} \exp\left\{ -\frac{i}{\hbar} \int_{t^\prime}^t ds \, H_0(s) \right\}  .
\end{equation}

%%%%%%%%%%%%%%%%%%%%%%%
\subsection{Unitary transformation}
Since the electrons in the junction leads and the electromagnetic environment are decoupled for vanishing tunneling, the time evolution operator (\ref{U0}) factorizes according to
\begin{equation}\label{Ufac}
U_0(t,t^{\prime})=U_{env}(t,t^{\prime}) \, e^{-\frac{i}{\hbar}H_{el}(t-t^{\prime})}
\end{equation}
into the time evolution operator 
\begin{equation}\label{Uenv}
U_{env}(t,t^{\prime}) = \mathcal{T} \exp\left\{ -\frac{i}{\hbar} \int_{t^\prime}^t ds \, H_{env}(s) \right\}  
\end{equation}
of the electrodynamic environment and the time evolution operator of the quasiparticles. 
The time dependence of the environmental Hamiltonian $H_{env}(t)$ arising from the applied voltage can be removed by a unitary transformation\cite{Grabert_2015}
\begin{eqnarray}\label{Lambda}
\Lambda(t)&=&\exp\left\{ \frac{i}{e}\left[\bar\varphi(t)Q +\sum_n\bar \varphi_n(t)Q_n  \right]\right\} \\ \nonumber && \times
\exp\left\{-\frac{i\hbar}{e^2}\left[C\dot{\bar\varphi}(t)\varphi+\sum_n C_n\dot{\bar\varphi}_n(t)\varphi_n\right]\right\}
\end{eqnarray}
where the phase $\bar \varphi(t)$ is determined by
\begin{equation}\label{eomphires2}
C\ddot{\bar\varphi}(t) + \int_0^t ds\, Y(t-s)\left[\dot{\bar\varphi}(s)-\dot\varphi_{ext}(s)\right] =0
\end{equation}
with the temporal response function
\begin{equation}\label{Yoft}
Y(t) = \sum_n \frac{1}{L_n} \cos(\omega_nt)
\end{equation}\noindent
of the electromagnetic environment which is the Fourier transform of the admittance
\begin{equation}\label{Yofom}
Y(\omega)=1/Z(\omega)=\int_0^{\infty}dt\, Y(t)\, e^{i\omega t} .
\end{equation}
The phases $\bar\varphi_n(t)$ are given by
\begin{equation}\nonumber
\bar\varphi_n(t)= \omega_n\int_0^t ds\, \sin\left[\omega_n(t-s)\right]\left[\bar\varphi(s)-\varphi_{ext}(s)\right] .
\end{equation}

Under the unitary transformation (\ref{Lambda}) the environmental Hamiltonian becomes\cite{Grabert_2015}
\begin{eqnarray}\label{trafHenv}\nonumber
\hat H_{env}(t)&=& \Lambda(t) H_{env}(t) \,\Lambda^{\dagger}(t) + i \hbar \frac{\partial \Lambda(t)}{\partial t}\Lambda^{\dagger}(t)  \\ 
&=& H^0_{env} + G(t)
\end{eqnarray}
where
\begin{equation}\label{Henv0}
H^0_{env} =\frac{Q^2}{2C}
+\sum_n \bigg[\frac{Q_n^2}{2C_n} 
+\frac{1}{2L_n}\left(\frac{\hbar}{e}\right)^2 \left(\varphi-\varphi_n\right)^2  \bigg] 
\end{equation}
is the time-independent Hamiltonian of the electromagnetic environment in the absence of driving. $G(t)$ is a function of time which gives rise to a phase factor in the time evolution operator \cite{Grabert_2015} dropping out in the Heisenberg representation of operators employed in the sequel.

Introducing the Heisenberg operators of the undriven electromagnetic environment
$
\tilde A(t) =  e^{\frac{i}{\hbar}H_{env}^0t} \, A\, e^{-\frac{i}{\hbar}H_{env}^0t}
$,
the phase operator in the interaction representation may be written as\cite{Grabert_2015}
\begin{equation}\label{phicheck2}
\check \varphi(t) = \tilde \varphi(t) +\bar \varphi(t) ,
\end{equation}
and for the interaction representation of the current operator (\ref{Ienv}) one finds
\begin{equation}\label{Ienvcheck}
\check I_{env}(t) =  \tilde I_{env}(t)  + \bar I_{env}(t)
\end{equation}
where
\begin{equation}\label{tildeIenv}
\tilde I_{env}(t) =\frac{\hbar}{e}C\ddot{\tilde\varphi}(t)
\end{equation}
is the current operator in the absence of driving and tunneling and
\begin{equation}\label{barIenv}
\bar I_{env}(t)  =\frac{\hbar}{e}C\ddot{\bar\varphi}(t)
\end{equation}
is the average displacement current flowing in a driven circuit in the absence of tunneling.\cite{Grabert_2015}
Hence, by virtue of the unitary transformation (\ref{Lambda}) all Heisenberg operators in the interaction representation on the rigth-hand side of Eq.~(\ref{IenvHeis}) can be expressed in terms of Heisenberg operators of the undriven system.

%%%%%%%%%%%%%%%%%%%%%
%%%%%%%%%%%%%%%%%%%%%
%%%%%%%%%%%%%%%%%%%%%
\section{Current autocorrelation function}\label{sec:three}
We are interested in properties of the system under quasi-stationary conditions, when the driving voltage $V_{ext}(t)$ has been acting for a long time.  Letting the initial time when driving starts tend to $-\infty$, one obtains from Eq.~(\ref{IenvHeis}) for the current operator
\begin{eqnarray}\label{IenvHeis2}\nonumber
&&I_{env}(t) = \check I_{env}(t)   \\ \nonumber
&&\quad
+\frac{i}{\hbar} \int_{0}^{\infty} du\,   \left[\check\Theta(t-u)\, e^{-i\check\varphi(t-u)}+ \hbox{H.c.}\, ,  \check I_{env}  (t)\right] \\ 
&&\quad -\frac{1}{\hbar^2}\int_{0}^{\infty} du \int_{0}^{\infty} dv\,   \Big[\check\Theta(t-u-v)\, e^{-i\check\varphi(t-u-v)}  \\ \nonumber
&&\quad\qquad  +\, \hbox{H.c.}\, ,  \Big[\check\Theta(t-u)\, e^{-i\check\varphi(t-u)}+ \hbox{H.c.}\, ,  \check I_{env}  (t)\Big] \Big] .
\end{eqnarray}
The average current in second order in $H_T$ resulting from this equation has been evaluated previously.\cite{Grabert_2015} Here we now determine the current autocorrelation function.

%%%%%%%%%%%%%%%%%%%%%%%
\subsection{Current autocorrelation function for weak tunneling}
The (non-symmetrized) current autocorrelation function is defined by
\begin{equation}
C(t,t^{\prime})= \langle  I_{env}(t) I_{env}(t^{\prime}) \rangle 
- \langle  I_{env}(t)\rangle\langle I_{env}(t^{\prime}) \rangle .
\end{equation}
This quantity is now determined up to terms of second order in the tunneling Hamiltonian. The result can be expressed entirely in terms of quantities known from the standard DCB theory.\cite{Ingold_1992} The outcome of the calculation for arbitrary time-dependent voltages is given in Eq.~(\ref{Ienvcorr6}) below, while the result for a superposition of a constant and a sinusoidal voltage drive is given in Eq.~(\ref{Ienvcorr8}).

It is advantageous to introduce the lag time $s$, the time difference between the two points in time 
$t$ and $t^{\prime}$. Inserting then the expansion (\ref{IenvHeis2}) of the Heisenberg operator $I_{env}(t)$ up to second order in the tunneling Hamiltonian $H_T$, we obtain for $C(t+s,t)$ a somewhat lengthy expression which is, however, readily written and given explicitly in Eq.~(\ref{Ienvcorr}) in App.~\ref{app:A}. All terms of the perturbative expansion refer to the unperturbed system with Hamiltonian $H_0=H_{el}+H_{env}$. Since the quasiparticle Hamiltonian $H_{el}$ and the environmental Hamiltonian $H_{env}$ commute, all terms of the expansion can be factorized into a quasiparticle average and an average over the electromagnetic environment. This is performed in App.~\ref{app:A} and the quasiparticle averages are expressed in terms of the correlator 
\begin{equation}\label{alpha}
\alpha(t)=\langle\check\Theta(t)\check\Theta^{\dagger}(0)\rangle=\langle\check\Theta^{\dagger}(t)\check\Theta(0)\rangle
\end{equation}
of tunneling operators. 

Furthermore, one inserts the decomposition (\ref{Ienvcheck}) of $\check I_{env}(t)$, the decomposition (\ref{phicheck2}) of $\check\varphi(t)$ as well as  the representation (\ref{tildeIenv}) of $\tilde I_{env}(t)$. As outlined in App.~\ref{app:A}, one then obtains for the current autocorrelation function the result (\ref{Ienvcorr3}) where all averages over the electromagnetic environment are expressed in terms of averages over the phase operator $\tilde\varphi(t)$ of the undriven circuit in the absence of tunneling. These averages can be expressed through the phase-phase correlation function
\begin{equation}\label{Joft}
J(t)=\left\langle  \left[ \tilde\varphi(t)- \tilde\varphi(0) \right] \tilde\varphi(0)  \right\rangle
\end{equation}
introduced in the standard $P(E)$ theory.\cite{Devoret_1990,Girvin_1990,Grabert_1991,Ingold_1992} The function $J(t)$ is given by
\begin{eqnarray}\label{Joft2}
J(t) &=& 2 \int_0^{\infty} \frac{d\omega}{\omega}\frac{Z^{\prime}_t(\omega)}{R_K}\\ \nonumber
&&\quad\times \left\{
\coth\left(\frac{1}{2}\beta\hbar\omega\right)\left [\cos(\omega t) - 1\right] - i \sin(\omega t)\right\}
\end{eqnarray}
where $Z^{\prime}_t(\omega)$ is the real part of the total impedance of the electromagnetic environment
\begin{equation}\label{Zt}
Z_t(\omega)=\frac{1}{Y(\omega)-i\omega C}
\end{equation}
and $R_K=h/e^2$ is the resistance quantum.

App.~\ref{app:A} presents the evaluation of  averages over the electromagnetic environment in terms of the phase-phase correlation function $J(t)$. Afterwards, 
the current autocorrelation function takes the form
\begin{eqnarray}\label{Ienvcorr6}
&&C(t+s,t)  =   \frac{\hbar^2C^2}{e^2}\ddddot J(s)
\\ \nonumber
&& \quad +\frac{2iC^2}{e^2}\int_{0}^{\infty} du \int_{0}^{\infty} dv\, \left(\alpha(v) 
 e^{J(v)} -\hbox{c.c.}\right) \\ \nonumber
&&\quad\quad\times\ddot J^{\prime\prime}(u)\left[\ddot J(s+u)-\ddot J(s+u+v)\right]\\ \nonumber
&&\quad\quad\times  \left( e^{i\bar\varphi(t-u)} e^{-i\bar\varphi(t-u-v)} +\hbox{c.c.}\right)
\\ \nonumber
&&\quad  +\frac{2iC^2}{e^2}\int_{0}^{\infty} du \int_{0}^{\infty} dv\, \left(\alpha(v)  e^{J(v)}-\hbox{c.c.}\right) \\ \nonumber
&&\quad\quad\times\ddot J^{\prime\prime}(u)  \left[\ddot J(s-u)-\ddot J(s-u-v) \right] \\ \nonumber
&&\quad\quad\times \left(e^{i\bar\varphi(t+s-u)} e^{-i\bar\varphi(t+s-u-v)} +\hbox{c.c.}\right)
\\ \nonumber
&&
\quad+\frac{2iC^2}{e^2} \int_{0}^{\infty} du\,    \int_{0}^{\infty} dv\,  \alpha(s-u+v) \,e^{J(s-u+v)}\\ \nonumber
&&\quad \quad\times\ddot J^{\prime\prime}(u)\left[\ddot J(-v)-\ddot J(v)\right] \\ \nonumber
&&\quad \quad\times\left( e^{-i\bar\varphi(t+s-u)}  e^{i\bar\varphi(t-v)} +\hbox{c.c.}\right)
\end{eqnarray}
which is written fully in terms of three quantities. The correlation function of tunneling operators $\alpha(t)$ defined in Eq.~(\ref{alpha}) and the phase autocorrelation function $J(t)$ defined in Eq.~(\ref{Joft}). These quantities characterize the quasiparticles and the electromagnetic environment, respectively, in the absence of tunneling and driving. Thirdly, the result (\ref{Ienvcorr6}) depends on the phase $\bar\varphi(t)$ determined by Eq.~(\ref{eomphires2}). This latter quantity describes the dependence on the external voltage.
So far the results are valid for any form of the driving potential $V_{ext}(t)$.

%%%%%%%%%%%%%%%%%%%
\subsection{Correlation function for periodic driving}
In the following, we focus on the important case of a circuit driven by the voltage
\begin{equation}\label{Vext}
V_{ext}(t)=V_{dc} + V_{ac}\cos\left(\Omega t\right)
\end{equation}
composed of a dc voltage $V_{dc}$ and a sinusoidal ac voltage $V_{ac}$ of frequency $\Omega$. 
In the absence of tunneling the junction charge then reads\cite{Grabert_2015} 
\begin{equation}\label{Qav}
\langle \check Q(t) \rangle = CV_{dc} +{\rm Re} \left\{\frac{  CY(\Omega)}{
Y(\Omega)- i\Omega C}V_{ac}\,e^{-i\Omega t}\right\}
\end{equation}
which oscillates with an amplitude proportional to $V_{ac}$ about
the time-averaged charge $CV_{dc}$. Using the polar decomposition 
\begin{equation}\label{polarde}
 \frac{Y(\omega)}{Y(\omega)-i\omega C}=\frac{Z_t(\omega)}{Z(\omega)} =\Xi(\omega) \, e^{i\eta(\omega)}
\end{equation}\noindent
of the admittance ratio into modulus $\Xi$ and phase $\eta$, Eq.~(\ref{Qav}) simply reads 
\begin{equation}
\langle \check Q(t) \rangle = C [V_{dc} + \Xi\,V_{ac}\cos(\Omega t -\eta)] .
\end{equation} 
Here and in the following we denote by $\Xi$ and $\eta$ without argument the functions $\Xi(\omega)$ and $\eta(\omega)$ for $\omega=\Omega$. Below, we will employ these functions also for other frequencies which are then always indicated explicitly as arguments.

Furthermore, one can show that the phase $\bar\varphi(t)$ introduced in Eqs.~(\ref{Lambda}) and (\ref{eomphires2}) coincides with the average phase $\langle \check\varphi(t) \rangle$ across the junction in the absence of tunneling and is given by \cite{Grabert_2015}
\begin{equation}
\bar\varphi(t) = \frac{e}{\hbar}V_{dc}t + a\sin(\Omega t -\eta)
\end{equation}
where we have introduced the dimensionless ac amplitude
\begin{equation}
a=\frac{e\,\Xi\,V_{ac}}{\hbar\Omega} .
\end{equation}
We now make use of
\begin{eqnarray}
&&\bar\varphi(t)-\bar\varphi(t-s)=\frac{e}{\hbar}V_{dc}s \\ \nonumber
&&\qquad+a\sin(\Omega t -\eta) - a\sin[\Omega (t-s) -\eta] .
\end{eqnarray}
Inserting this into the result (\ref{Ienvcorr6}), the current autocorrelation function takes the form
\begin{eqnarray}\label{Ienvcorr8}
&&C(t+s,t)  =   \frac{\hbar^2C^2}{e^2}\ddddot J(s)
\\ \nonumber
&& \quad +\frac{2iC^2}{e^2}\int_{0}^{\infty} du \int_{0}^{\infty} dv\, \left(\alpha(v) 
 e^{J(v)} -\hbox{c.c.}\right) \\ \nonumber
&&\quad\quad\times\ddot J^{\prime\prime}(u)\left[\ddot J(s+u)-\ddot J(s+u+v)\right] \\ \nonumber
&&\quad\quad\times\!\left( e^{\frac{i}{\hbar}eV_{dc}v}e^{ia\sin[\Omega(t-u)-\eta]}e^{-ia\sin[\Omega(t-u-v)-\eta]} +\hbox{c.c.}\right)
\\ \nonumber
&&  \quad+\frac{2iC^2}{e^2}\int_{0}^{\infty} du \int_{0}^{\infty} dv\,\left(\alpha(v)  e^{J(v)}-\hbox{c.c.}\right)  \\ \nonumber
&& \quad\quad\times\ddot J^{\prime\prime}(u)\left[\ddot J(s-u)-\ddot J(s-u-v) \right]  \\ \nonumber
&& \quad\quad\times\Big( e^{\frac{i}{\hbar}eV_{dc}v}\, e^{ia\sin[\Omega(t+s-u)-\eta]}\,\\ \nonumber
&&\quad\quad\times e^{-ia\sin[\Omega(t+s-u-v)-\eta]} +\hbox{c.c.}\Big)
\\ \nonumber
&&\quad
+\frac{4C^2}{e^2} \int_{0}^{\infty} du\,    \int_{0}^{\infty} dv\,  \alpha(s-u+v)\\ \nonumber
&&\quad\quad\times e^{J(s-u+v)}\ddot J^{\prime\prime}(u)\,\ddot J^{\prime\prime}(v) \Big( e^{\frac{i}{\hbar}eV_{dc}(s-u+v)}\,\\ \nonumber
&&\quad\quad\times e^{ia\sin[\Omega(t+s-u)-\eta]}  \, e^{-ia\sin[\Omega(t-v)-\eta]} +\hbox{c.c.}\Big) .
\end{eqnarray}
To proceed it is advantageous to turn to the spectral density of current fluctuations. 

%%%%%%%%%%%%%%%%%%%%%%%%
%%%%%%%%%%%%%%%%%%%%%%%%
%%%%%%%%%%%%%%%%%%%%%%%%
\section{Current spectral function}\label{sec:four}
The spectrum of current fluctuations is experimentally particularly relevant. Accordingly, we continue with the analysis in Fourier space. We will show that the spectrum is composed of three parts given in Eqs.~(\ref{SNomega2}), (\ref{STomega5}) and (\ref{SNTomega4}) below.
%%%%%%%%%%%%%%%%%%%%%%%%
\subsection{Fourier coefficients of current autocorrelation function}
The correlation function (\ref{Ienvcorr8}) has the periodicity
\begin{equation}
C(t+{2\pi}/{\Omega}+s,t+{2\pi}/{\Omega})= C(t+s,t)
\end{equation}
and may thus be written as a Fourier series
\begin{equation}
C(t+s,t)=\sum_{n=-\infty}^{\infty} C_n(s)\, e^{-in(\Omega t-\eta)}
\end{equation}
with coefficients $C_n(s)$ that are functions of the lag time $s$.	
To extract these coefficients, we employ the Jacobi-Anger expansion\cite{Abramowitz} 
\begin{equation}
e^{ia\sin(\Omega t)}= \sum_{k=-\infty}^{\infty} J_k(a)\, e^{ik\Omega t}
\end{equation}
with the Bessel functions of the first kind $J_k(z)$. Inserting this series expansion
into Eq.~(\ref{Ienvcorr8}), we obtain for the current autocorrelation function the expression
\begin{eqnarray}\label{Ienvcorr10}
&&C(t+s,t)  =   \frac{\hbar^2C^2}{e^2}\ddddot J(s)
\\ \nonumber
&& \quad +\frac{2iC^2}{e^2}\int_{0}^{\infty} du \, \ddot J^{\prime\prime}(u)\int_{0}^{\infty} dv\, \left(\alpha(v) 
 e^{J(v)} -\hbox{c.c.}\right) \\ \nonumber
&&\quad\quad\times\left[\ddot J(s+u)-\ddot J(s+u+v)\right] \sum_{k,l=-\infty}^{\infty}J_k(a)\, J_l(a)\\ \nonumber
&&\quad\quad\times
\left( e^{\frac{i}{\hbar}eV_{dc}v}e^{ik[\Omega(t-u)-\eta]}e^{-il[\Omega(t-u-v)-\eta]} +\hbox{c.c.}\right)
\\ \nonumber
&&  \quad+\frac{2iC^2}{e^2}\int_{0}^{\infty} du\, \ddot J^{\prime\prime}(u) \int_{0}^{\infty} dv\,\left(\alpha(v)  e^{J(v)}-\hbox{c.c.}\right)  \\ \nonumber
&& \quad\quad\times\left[\ddot J(s-u)-\ddot J(s-u-v) \right]   \sum_{k,l=-\infty}^{\infty}J_k(a)\, J_l(a) \\ \nonumber
&& \quad\quad\times
\left(\! e^{\frac{i}{\hbar}eV_{dc}v}\, e^{ik[\Omega(t+s-u)-\eta]}\, e^{-il[\Omega(t+s-u-v)-\eta]} \!+\! \hbox{c.c.}\right)
\\ \nonumber
&&\quad
+\frac{4C^2}{e^2} \int_{0}^{\infty} du\,  \ddot J^{\prime\prime}(u)  \int_{0}^{\infty} dv\, \ddot J^{\prime\prime}(v) \,\\ \nonumber
&&\quad\quad\times \alpha(s-u+v)\,e^{J(s-u+v)}\sum_{k,l=-\infty}^{\infty}J_k(a)\, J_l(a)\\ \nonumber
&&\quad\quad\times\!
\left(\! e^{\frac{i}{\hbar}eV_{dc}(s-u+v)}e^{ik[\Omega(t+s-u)-\eta]} e^{-il[\Omega(t-v)-\eta]}\!  +\! \hbox{c.c.}\right) .
\end{eqnarray}
From this result we can now easily extract the Fourier coefficients $C_n(s)$. In particular, we obtain for the most important coefficient $C_0(s)$, which corresponds to the average of the correlator $C(t+s,t)$ over time $t$, the result
\begin{eqnarray}\label{C0ofs}
&&C_0(s)   =   \frac{\hbar^2C^2}{e^2}\ddddot J(s)
\\ \nonumber
&& \quad +\frac{2iC^2}{e^2}\int_{0}^{\infty} du \, \ddot J^{\prime\prime}(u)\int_{0}^{\infty} dv\, \left(\alpha(v) 
 e^{J(v)} -\hbox{c.c.}\right) \\ \nonumber
&&\quad\quad\times\left[\ddot J(s+u)-\ddot J(s+u+v)\right]  \\ \nonumber
&&\quad\quad\times\sum_{k=-\infty}^{\infty}J_k(a)^2
\left( e^{\frac{i}{\hbar}eV_{dc}v}e^{ik\Omega v} +\hbox{c.c.}\right)
\\ \nonumber
&&  \quad+\frac{2iC^2}{e^2}\int_{0}^{\infty} du\, \ddot J^{\prime\prime}(u) \int_{0}^{\infty} dv\,\left(\alpha(v)  e^{J(v)}-\hbox{c.c.}\right)  \\ \nonumber
&& \quad\quad\times\left[\ddot J(s-u)-\ddot J(s-u-v) \right]  \\ \nonumber
&& \quad\quad\times \sum_{k=-\infty}^{\infty}J_k(a)^2
\left( e^{\frac{i}{\hbar}eV_{dc}v}\, e^{ik\Omega v}+\hbox{c.c.}\right)
\\ \nonumber
&&\quad
+\frac{4C^2}{e^2} \int_{0}^{\infty} du\,  \ddot J^{\prime\prime}(u)  \int_{0}^{\infty} dv\, \ddot J^{\prime\prime}(v) \\ \nonumber
&&\quad\quad\times\alpha(s-u+v)\,e^{J(s-u+v)} \\ \nonumber
&&\quad\quad\times \sum_{k=-\infty}^{\infty}J_k(a)^2
\left( e^{\frac{i}{\hbar}eV_{dc}(s-u+v)}\, e^{ik\Omega(s-u+v)}   +\hbox{c.c.}\right) .
\end{eqnarray}
We now switch to Fourier space also with respect to the lag time $s$.

%%%%%%%%%%%%%%%%%%%%%%
\subsection{Spectral function}
The Fourier coefficient $C_0(s)$ of the current correlator can be written as a Fourier integral
\begin{equation}
C_0(s)=\int_{-\infty}^{\infty}\frac{d\omega}{2\pi}\, S(\omega)\, e^{-i\omega s}
\end{equation}
with the spectral function
\begin{equation}
S(\omega)=\int_{-\infty}^{\infty} ds\, C_0(s)\, e^{i\omega s} .
\end{equation}
To determine $S(\omega)$ it is advantageous to transform the result (\ref{C0ofs}) for $C_0(s)$  in such a way that the dependence on $s$ arises via exponential factors of the form $\exp(i\nu s)$. To this purpose we make use of Eq.~(\ref{Joft2}) which gives
\begin{eqnarray}\label{ddJoft2}
\ddot J(t) &=&- 2 \int_0^{\infty} d\omega\,\omega\,\frac{Z^{\prime}_t(\omega)}{R_K}\\ \nonumber
&&\quad\times \left\{
\coth\left(\frac{1}{2}\beta\hbar\omega\right) \cos(\omega t)  - i \sin(\omega t)\right\} .
\end{eqnarray}
This may be transformed to read as
\begin{equation}\label{ddJoft4}
\ddot J(t) =- \frac{e^2}{\pi\hbar} \int_{-\infty}^{\infty} d\omega\,\frac{\omega\, Z^{\prime}_t(\omega)\,e^{-i\omega t}}{1- e^{
-\beta\hbar\omega}} 
\end{equation}
where we have made use of the symmetry 
\begin{equation}
Z^{\prime}_t(\omega) =Z^{\prime}_t(-\omega)
\end{equation}
following from Eq.~(\ref{Zt}) and the relation
\begin{equation}
\frac{1}{1-e^{-x}}=\tfrac{1}{2}\left[1+\coth\left(\tfrac{x}{2}\right)\right] .
\end{equation}
It is convenient to introduce also the function
\begin{equation}
K(s) = -\ddot J(s) = \langle \dot {\tilde\varphi}(t+s) \dot {\tilde\varphi}(t)\rangle
\end{equation}
which is proportional to the charge autocorrelation function in the absence of tunneling and driving, and its Fourier representation
\begin{equation}\label{Koft}
K(s)=\int_{-\infty}^{\infty}\frac{d\omega}{2\pi}\, S_K(\omega)\, e^{-i\omega s} .
\end{equation}
From Eq.~(\ref{ddJoft4}) we obtain for the spectrum
\begin{equation}\label{SKofom}
S_K(\omega)=  \frac{2 e^2}{\hbar}  \frac{\omega\,Z^{\prime}_t(\omega)}{1- e^{
-\beta\hbar\omega}} 
\end{equation}
which is real, implying $K(-s)=K(s)^*$, and which has the symmetry
\begin{equation}\label{symSK}
S_K(-\omega)=e^{-\beta\hbar\omega}S_K(\omega) .
\end{equation}

To uncover the dependence of $C_0(s)$ on $s$, we also need a Fourier representation of the function $\alpha(t)\, e^{J(t)}$. To this purpose we introduce the familiar $P(E)$-function of the DCB theory\cite{Devoret_1990,Girvin_1990,Grabert_1991,Ingold_1992}
\begin{equation}\label{PofE}
P(E)=\frac{1}{2\pi\hbar}\int_{-\infty}^{\infty}dt\, e^{J(t)+\frac{i}{\hbar}Et}
\end{equation}
which gives the probability that a tunneling transition is associated with an exchange of the energy $E$ with the electromagnetic environment. Now, Eq.~(\ref{PofE}) implies
\begin{equation}\label{FFeJ}
e^{J(t)}= \hbar \int_{-\infty}^{\infty}\,d\omega\, P(\hbar\omega)\, e^{-i\omega t} .
\end{equation}
Furthermore, for tunnel junctions in the wide band limit, the correlator of tunneling operators $\alpha(t)$ is given by\cite{Ingold_1992,Grabert_2015}
\begin{equation}\label{ThetaTheta4}
\alpha(t) = \frac{1}{2\pi}\frac{\hbar\,G_T}{e^2}\int_{-\infty}^{\infty} dE\, \frac{E\, e^{-\frac{i}{\hbar}E t} }{1-e^{-\beta E}} 
\end{equation}
where $G_T$ is the tunneling conductance of the junction.
Combining Eqs.~(\ref{FFeJ}) and (\ref{ThetaTheta4}) one obtains
\begin{eqnarray}\label{FFalphaJ}
\alpha(t)\, e^{J(t)}&=&\frac{2\pi\hbar^4G_T}{e^2}\int_{-\infty}^{\infty}  \frac{d\omega}{2\pi}\int_{-\infty}^{\infty}  \frac{d\omega^{\prime}}{2\pi}\\ \nonumber
&& \times \frac{\omega\, P(\hbar\omega^{\prime})}{1-e^{-\beta \hbar\omega}}  e^{-i(\omega+\omega^{\prime}) t}  .
\end{eqnarray}
We write this result in the form
\begin{equation}
\alpha(t)\, e^{J(t)}=\int_{-\infty}^{\infty}  \frac{d\omega}{2\pi} F(\omega)\, e^{-i\omega  t} 
\end{equation}
with the spectrum
\begin{equation}\label{Fomega}
F(\omega)= \frac{\hbar^4G_T}{e^2} \int_{-\infty}^{\infty}d\omega^{\prime}\, \frac{(\omega-\omega^{\prime})\, P(\hbar\omega^{\prime})}{1-e^{-\beta \hbar(\omega-\omega^{\prime})}}
\end{equation}
which obeys the symmetry
\begin{equation}
F(-\omega)=e^{-\beta\hbar\omega}F(\omega) .
\end{equation}
Using the Fourier representations (\ref{Koft}) and (\ref{FFalphaJ}), the expression (\ref{C0ofs}) for the function $C_0(s)$ can be transformed in such a way that  the spectral function $S(\omega)$ can easily be extracted. This is shown explicitly in App.~\ref{app:B}. The resulting spectral function is found to be the sum of three terms
\begin{equation}\label{Somegasum}
S(\omega)=   S_N(\omega)+ S_T(\omega) +  S_{NT}(\omega)
\end{equation}
where
\begin{equation}\label{SNomega}
S_N(\omega)= \frac{\hbar^2C^2}{e^2}  \, \omega^2\, S_K(\omega) ,
\end{equation}
\begin{eqnarray}\label{STomega}\nonumber
&&S_T(\omega) = 
-\frac{ C^2}{e^2} \int_{0}^{\infty}\!\! du\,  
\int_{-\infty}^{\infty} \frac{d \mu}{2\pi} \left(1-e^{-\beta\hbar\mu}\right) S_K(\mu) e^{-i\mu u}  \\  \nonumber
&& \quad\times \int_{0}^{\infty}\! dv
\int_{-\infty}^{\infty} \frac{d \nu}{2\pi} \left(1-e^{-\beta\hbar\nu}\right) S_K(\nu)\,e^{-i\nu v} e^{i\omega(u-v)}
\\ \nonumber && \quad \times\sum_{k=-\infty}^{\infty}J_k(a)^2\,
\big[ F(\omega+eV_{dc}/\hbar+k\Omega)\\ 
&&\qquad\qquad\qquad+  F(\omega-eV_{dc}/\hbar-k\Omega)\big]
\end{eqnarray}
and
\begin{eqnarray}\label{SNTomega}\nonumber
&&S_{NT}(\omega) =  \frac{C^2}{e^2}\int_{0}^{\infty}\!\! du \int_{-\infty}^{\infty} \frac{d \mu}{2\pi}\, \left(1-e^{-\beta\hbar\mu}\right) S_K(\mu) e^{-i\mu u}\\  \nonumber
&& \quad
 \times \int_{0}^{\infty}\! dv\, \left(\alpha(v) 
 e^{J(v)} -\hbox{c.c.}\right)  S_K(\omega)\, e^{-i\omega u}\left[1-e^{-i\omega v}\right] \\  
&& \quad
 \times  \sum_{k=-\infty}^{\infty}J_k(a)^2
\left( e^{\frac{i}{\hbar}eV_{dc}v}e^{ik\Omega v} +\hbox{c.c.}\right)\ + \hbox{c.c.}\ .
\end{eqnarray}
These three contributions to $S(\omega)$ are now evaluated further.

%%%%%%%%%%%%%%%
\subsection{Evaluation of \bm{$S_N(\omega)$}}
The contribution $S_N(\omega)$ to the spectral function $S(\omega)$ may be written as 
\begin{equation}\label{SNomega2}
S_N(\omega)=  \frac{2\hbar\,C^2 \omega^3 Z^{\prime}_t(\omega)}{1- e^{
-\beta\hbar\omega}} 
\end{equation} 
where we have inserted the explicit form (\ref{SKofom}) of the spectrum $S_K(\omega)$ into Eq.~(\ref{SNomega}).
Now, the real part $Z^{\prime}_t(\omega)$ of the total impedance (\ref{Zt}) may be written as
\begin{eqnarray}
Z^{\prime}_t(\omega)&=& \vert Z_t(\omega)\vert^2 Y^{\prime}(\omega) =
\left\vert\frac{Z_t(\omega)}{ Z(\omega)}\right\vert^2 Z^{\prime}(\omega)\\ \nonumber
&=&\Xi(\omega)^2  Z^{\prime}(\omega)
\end{eqnarray}
where we have made use of the polar decomposition (\ref{polarde}) to obtain the last expression. This allows us to write the result (\ref{SNomega2}) in the form
\begin{equation}\label{SNomega3}
S_N(\omega)= 
\left[\Xi(\omega)\omega Z^{\prime}(\omega)  C\right]^2   S_{I_NI_N}(\omega)
\end{equation}
where
\begin{equation}
S_{I_NI_N}(\omega)=\frac{2\hbar\omega }{1- e^{
-\beta\hbar\omega}} \frac{1}{Z^{\prime}(\omega)}
\end{equation}
is the spectrum of Johnson-Nyquist current noise\cite{Johnson_1932,Nyquist_1932,Callen_1951} generated by the environmental impedance $Z(\omega)$. We thus see that the contribution $S_N(\omega)$ to the spectral function $S(\omega)$ is the current noise in the outer circuit arising from the Johnson-Nyquist noise of the environmental impedance. The dimensionless factor  
$
 \left[\Xi(\omega) \omega Z^{\prime}(\omega)  C\right]^2
$
in Eq.~(\ref{SNomega3}) can be understood from circuit theory.

%%%%%%%%%%%%%%%
\subsection{Evaluation of \bm{$S_T(\omega)$}}
Next, we turn to the contribution $S_T(\omega)$ to the spectral function $S(\omega)$.
We start by noting that the average current $I_{dc}(V_{dc})$ in the presence of a dc voltage $V_{dc}$ only may be written as\cite{Ingold_1992}
\begin{equation}\label{Idc3}
I_{dc}(V)=e\left[\Gamma(V)-\Gamma(-V)\right]
\end{equation}
where
\begin{eqnarray}\label{rate}
\Gamma(V)&=&\frac{G_T}{e^2}\int_{-\infty}^{\infty} d E\, \frac{E\, P(eV-E)}{1-e^{-\beta E}}\\ \nonumber
&=&\frac{\hbar^2 G_T}{e^2}\int_{-\infty}^{\infty} d \omega^{\prime}\, \frac{(eV/\hbar-\omega^{\prime})\, P(\hbar\omega^{\prime})}{1-e^{-\beta (eV-\hbar\omega^{\prime})}}
\end{eqnarray}
is the electron tunneling rate in the presence of an applied dc voltage $V$. When we compare this with the definition (\ref{Fomega}) of the spectrum $F(\omega)$ introduced previously, we see that $F(\omega)$ may be written as 
\begin{equation}
F(\omega)=\hbar^2\Gamma(\hbar\omega/e) .
\end{equation}
Inserting this result into Eq.~(\ref{STomega}) we obtain
\begin{eqnarray}\nonumber\label{STomega2}
&&S_T(\omega)  =  -\frac{\hbar^2 C^2}{e^2} \!
\int_{0}^{\infty}\!\! du 
\int_{-\infty}^{\infty} \frac{d \mu}{2\pi} \left(1-e^{-\beta\hbar\mu}\right) S_K(\mu)
\\  \nonumber
&& \quad\times e^{-i\mu u} \int_{0}^{\infty}\!\! dv
\int_{-\infty}^{\infty} \frac{d \nu}{2\pi} \left(1-e^{-\beta\hbar\nu}\right) S_K(\nu)e^{-i\nu v}
\\  \nonumber
&& \quad \times \sum_{k=-\infty}^{\infty}J_k(a)^2
\big[\Gamma\left(V_{dc}+\hbar(k\Omega+\omega)/e\right)  e^{i\omega (u-v)} \\ 
&& \qquad
+ \Gamma\left(-V_{dc}-\hbar(k\Omega-\omega)/e\right)  e^{i\omega (u-v)} \big] .
\end{eqnarray}
Now, the rigth-hand side of this relation contains the integrals
\begin{equation}
\int_0^{\infty} du\, e^{-i\mu u}\, e^{i\omega u} =\pi\delta(\mu-\omega)-iP\frac{1}{\mu-\omega}
\end{equation}
and
\begin{equation}
\int_0^{\infty} dv\, e^{-i\nu v}\, e^{-i\omega v} =\pi\delta(\nu+\omega)-iP\frac{1}{\nu+\omega}
\end{equation}
where $P$ denotes the Cauchy principal value.  
When we apply these relations, Eq.~(\ref{STomega2}) takes the form
\begin{eqnarray}\label{STomega3}
&&S_T(\omega)  =  -\frac{\hbar^2 C^2}{e^2} 
\Bigg[\frac{1}{2}\left(1-e^{-\beta\hbar\omega}\right) S_K(\omega)\\ \nonumber
&&\qquad -iP\int_{-\infty}^{\infty}\, \frac{d \mu}{2\pi}\, \frac{\left(1-e^{-\beta\hbar\mu}\right) S_K(\mu)}{\mu-\omega}
\Bigg] \\  \nonumber
&& \quad \times 
\Bigg[\frac{1}{2}\left(1-e^{\beta\hbar\omega}\right) S_K(-\omega)  \\  \nonumber
&& \qquad -iP\int_{-\infty}^{\infty}\, \frac{d \nu}{2\pi}\,\frac{ \left(1-e^{-\beta\hbar\nu}\right) S_K(\nu)}{\nu+\omega}
\Bigg] \sum_{k=-\infty}^{\infty}J_k(a)^2 \\ \nonumber
&& \quad\times
\big[\Gamma\left(V_{dc}+\hbar(k\Omega+\omega)/e\right)  
+ \Gamma\left(-V_{dc}-\hbar(k\Omega-\omega)/e\right)   \big] .
\end{eqnarray}
This result contains the Cauchy principal value
\begin{equation}\label{HK}
H_K(\omega) = P \int_{-\infty}^{\infty}\, \frac{d \mu}{2\pi}\, \frac{\left(1-e^{-\beta\hbar\mu}\right) S_K(\mu)}{\mu-\omega}
\end{equation}
which is evaluated next.

%%%%%%%%%%%%%%%%%%%%%%%
\subsubsection{Evaluation of $H_K(\omega)$}
To determine $H_K(\omega)$ we first note that Eq.~(\ref{SKofom}) gives
\begin{equation}\label{SKofom2}
\left(1-e^{-\beta\hbar\omega}\right)S_K(\omega)=  \frac{2 e^2}{\hbar}\omega  
Z^{\prime}_t(\omega)
\end{equation}
so that
\begin{eqnarray}\label{HK2}
H_K(\omega) &=& \frac{2 e^2}{\hbar} P \int_{-\infty}^{\infty}\, \frac{d \mu}{2\pi}\, \frac{\mu  Z^{\prime}_t(\mu) }{\mu-\omega} \\ \nonumber
&=& \frac{2 e^2}{\hbar} \hbox{ Re}\left[ P \int_{-\infty}^{\infty}\, \frac{d \mu}{2\pi}\, 
\frac{\mu\, Z_t(\mu) }{\mu-\omega} \right] .
\end{eqnarray}
We now make use of
\begin{equation}
\frac{\mu}{\mu-\omega}=1+\frac{\omega}{\mu-\omega}
\end{equation}
which implies
\begin{eqnarray}
&&P \int_{-\infty}^{\infty}\, \frac{d \mu}{2\pi}\, \frac{\mu\, Z_t(\mu) }{\mu-\omega}\\ \nonumber
&&=
\int_{-\infty}^{\infty}\, \frac{d \mu}{2\pi}\,  Z_t(\mu) +\omega P \int_{-\infty}^{\infty}\, \frac{d \mu}{2\pi}\, \frac{Z_t(\mu) }{\mu-\omega} .
\end{eqnarray}
The total impedance $Z_t(\omega)$ obeys the sum rule\cite{Girvin_1990,Ingold_1992}
\begin{equation}
\int_{-\infty}^{\infty} d\omega\, Z_t(\omega) =\frac{\pi}{C}
\end{equation}
and the Kramers-Kronig relations imply
\begin{equation}
P\int_{-\infty}^{\infty}d\omega\, \frac{Z_t(\omega)}{\omega-\omega^{\prime}}=i\pi Z_t(\omega^{\prime}) .
\end{equation}
With the help of these relation we obtain 
\begin{equation}
P \int_{-\infty}^{\infty}\, \frac{d \mu}{2\pi}\, \frac{\mu \,Z_t(\mu) }{\mu-\omega}\\ \nonumber
=\frac{1}{2C} +\frac{i}{2}\omega Z_t(\omega) 
\end{equation}
which combines with Eq.~(\ref{HK2}) to give
\begin{equation}\label{HK3}
H_K(\omega)=\frac{e^2}{\hbar}\left[\frac{1}{C} -\omega\, Z_t^{\prime\prime}(\omega)\right]
\end{equation}
which has the symmetry
\begin{equation}
H_K(\omega)=H_K(-\omega) .
\end{equation}

%%%%%%%%%%%%%%%%%%%%%
\subsubsection{Evaluation of $S_T(\omega)$ continued}
Expressing the right-hand side of Eq.~(\ref{STomega3}) in terms of $H_K(\omega)$ and using then Eq.~(\ref{HK3}), we obtain 
\begin{eqnarray}\label{STomega4}
&&S_T(\omega)  =  
\bigg(\frac{\hbar^2 C^2}{ e^2} \left[\frac{1}{2}\left(1-e^{-\beta\hbar\omega}\right) S_K(\omega)\right]^2 \\ \nonumber
&&\quad
+e^2\left[1 -\omega\,C\, Z_t^{\prime\prime}(\omega)\right]^2
\bigg) \sum_{k=-\infty}^{\infty}J_k(a)^2
\\  \nonumber
&& \quad \times 
\left[\Gamma\left(V_{dc}+\hbar(k\Omega+\omega)/e\right)
+ \Gamma\left(-V_{dc}-\hbar(k\Omega-\omega)/e\right)  \right] .
\end{eqnarray}
Next, we note that\cite{Ingold_1992}
\begin{equation}
P(-E)=e^{-\beta E}P(E) .
\end{equation}
Hence, Eq.~(\ref{rate}) implies
\begin{equation}
\Gamma(-V)=e^{-\beta e V} \Gamma(V) .
\end{equation}
The current  (\ref{Idc3}) may thus be written in the form
\begin{equation}
I_{dc}(V)=e\left(1-e^{-\beta e V}\right)\Gamma(V) .
\end{equation}
Vice versa, the rate can be expressed as
\begin{equation}\label{rate2}
\Gamma(V)=\frac{I_{dc}(V)}{e\left(1-e^{-\beta eV}\right)}\,, \quad \Gamma(-V)=\frac{I_{dc}(V)}{e\left(e^{\beta eV}-1\right)} .
\end{equation}

Further, using the explicit form (\ref{SKofom}) of $S_K(\omega)$ we obtain
\begin{equation}
\frac{\hbar^2C^2}{e^2}\left[\frac{1}{2}\left(1-e^{-\beta\hbar\omega}\right) S_K(\omega)\right]^2 =
 e^2[\omega C  Z_t^{\prime}(\omega)]^2 .
\end{equation}
The first factor in Eq.~(\ref{STomega4}) may thus be written as
\begin{eqnarray}\label{factor2}\nonumber
&&
 \frac{\hbar^2 C^2}{ e^2} \left[\frac{1}{2}\left(1-e^{-\beta\hbar\omega}\right) S_K(\omega)\right]^2 \!
+e^2\left[1 -\omega\,C\, Z_t^{\prime\prime}(\omega)\right]^2
 \\ \nonumber
&&\quad=
e^2\left([\omega C  Z_t^{\prime}(\omega)]^2 + \left[1 -\omega C Z_t^{\prime\prime}(\omega)\right]^2\right)
\\ 
&&\quad=e^2\left\vert \frac{Z_t(\omega)}{Z(\omega)}\right\vert^2 =e^2\Xi(\omega)^2
\end{eqnarray}
where we have employed the form~(\ref{Zt}) of $Z_t(\omega)$ and the polar decomposition (\ref{polarde}).

Using now the relations (\ref{rate2}) and (\ref{factor2}), the result (\ref{STomega4}) can be cast in the form
\begin{eqnarray}\label{STomega5}
&&S_T(\omega)  =  e\Xi(\omega)^2
\sum_{k=-\infty}^{\infty}J_k(a)^2
\\  \nonumber
&& \times 
\bigg\{\frac{I_{dc}\left(V_{dc}+\hbar(k\Omega+\omega)/e\right)}{1-e^{-\beta\left[eV_{dc}+\hbar(k\Omega+\omega)\right]}}
+ \frac{I_{dc}\left(V_{dc}+\hbar(k\Omega-\omega)/e\right)}{e^{\beta\left[eV_{dc}+\hbar(k\Omega-\omega)\right]}-1}  \bigg\}
\end{eqnarray}
which expresses the contribution $S_T(\omega)$ to the spectral function in terms of the current-voltage relation for the average current $I_{dc}(V_{dc})$ in the absence of an ac driving voltage.

%%%%%%%%%%%%%%%%%%%%%%
\subsubsection{Relation of $S_T(\omega)$ to shot noise}

In the absence of an ac voltage, i.e., when $a=0$, all Bessel functions $J_k(0)=0$ except for $J_0(0)=1$, and the result (\ref{STomega5}) simplifies to read as
\begin{eqnarray}\label{STdc}
&&S_T(V_{dc},0,\omega)  =   e\Xi(\omega)^2
\\  \nonumber
&& \qquad\times 
\bigg\{\frac{I_{dc}\left(V_{dc}+\hbar\omega/e\right)}{1-e^{-\beta\left[eV_{dc}+\hbar\omega\right]}}
+ \frac{I_{dc}\left(V_{dc}-\hbar\omega/e\right)}{e^{\beta\left[eV_{dc}-\hbar\omega\right]}-1}  \bigg\}
\end{eqnarray}
where we have made the voltage dependence of the spectral function explicit using the notation $S(V_{dc},V_{ac},\omega)$ for the spectral function and its parts whenever appropriate.

The result (\ref{STdc}) may be written as\cite{Lee_1996} 
\begin{equation}\label{STdc2}
S_T(V_{dc},0,\omega)  =  \Xi(\omega)^2
S_{I_TI_T}(V_{dc},\omega)
\end{equation}
where
\begin{equation}
S_{I_TI_T}(V,\omega)=e\left[\frac{ I_{dc}(eV+ \hbar\omega)}{1-e^{-\beta(eV+\hbar\omega)}}
+\frac{ I_{dc}(eV- \hbar\omega)}{e^{\beta(eV-\hbar\omega)}-1}\right]
\end{equation}      
is the shot noise of the tunneling current\cite{Rovogin_1974,Lee_1996} caused by electron tunneling across a tunnel junction with applied dc voltage $V$.
This shows that the component $S_T(\omega)$ of the spectral function comes from the shot noise generated at the tunneling element. There is a frequency-dependent factor $ \Xi(\omega)^2$ in Eqs.~(\ref{STomega5}) -- (\ref{STdc2}) relating the noise of the tunneling current with the noise of the environmental current. This factor can also be derived from circuit theory.

When we compare the results (\ref{STomega5}) and (\ref{STdc}), we see that the contribution $S_T(V_{dc},V_{ac},\omega)$ to the spectral function $S(V_{dc},V_{ac},\omega)$ in the presence of dc and ac voltages is related to the corresponding contribution $S_T(V_{dc},0,\omega)$ in the presence of a dc voltage only by a photo-assisted tunneling relation
\begin{equation}\label{paT}
S_T(V_{dc},V_{ac},\omega)=\sum_{k=-\infty}^{\infty}J_k(a)^2\, S_T(V_{dc}+k\hbar\Omega/e,0,\omega)
\end{equation}
of the Tien-Gordon type.\cite{Tien_1963,Tucker_1985}. 

%%%%%%%%%%%%%%%
\subsection{Evaluation of \bm{$S_{NT}(\omega)$}}

Finally, we need to determine the contribution $S_{NT}(\omega)$ to the spectral function $S(\omega)$. We start by noting that the right-hand side of Eq.~(\ref{SNTomega}) contains the integral
\begin{equation}
\int_0^{\infty} du \,e^{-i\mu u}\,e^{-i\omega u}=\pi \delta(\mu+\omega)-iP\frac{1}{\mu+\omega} .
\end{equation}
Accordingly, we obtain from Eq.~(\ref{SNTomega})
\begin{eqnarray}\label{SNTomega2}
&&S_{NT}(\omega)  =   -\frac{C^2}{2e^2}\Bigg[\left(1-e^{-\beta\hbar\omega}\right) S_K(\omega)^2\\ \nonumber
&&\qquad +2i\,  S_K(\omega)\, P \int_{-\infty}^{\infty}\, \frac{d \mu}{2\pi}\,\frac{ \left(1-e^{-\beta\hbar\mu}\right) S_K(\mu)}{\mu+\omega}\Bigg]  \\  \nonumber
&& \qquad
 \times\!\sum_{k=-\infty}^{\infty}J_k(a)^2 \!\int_{0}^{\infty} dv\, \left(\alpha(v) 
 e^{J(v)} -\hbox{c.c.}\right)\!\left[1-e^{-i\omega v}\right] \\ \nonumber
 &&\qquad\times
\left( e^{\frac{i}{\hbar}eV_{dc}v}e^{ik\Omega v} +\hbox{c.c.}\right)  \ +\hbox{c.c.}
\end{eqnarray}
where we have exploited the symmetry (\ref{symSK}) of $S_K(\omega)$.
Now, the result (\ref{SNTomega2}) contains the Cauchy principal value $H_K(\omega)$ introduced in Eq.~(\ref{HK})
and integrals of the form
\begin{equation}\label{intv}
L(\omega)=\int_{0}^{\infty} dv\, \left(\alpha(v)  e^{J(v)}-\hbox{c.c.}\right)e^{i\omega v}
\end{equation}
which are evaluated next.

%%%%%%%%%%%%%%%%%%%%
\subsubsection{Evaluation of $L(\omega)$}
To evaluate the integral $L(\omega)$ defined in Eq.~(\ref{intv}) we write it in the form
\begin{equation}\label{intv2}
 L(\omega)=  \int_{0}^{\infty}\! dv\, \alpha(v)  e^{J(v)}\,e^{i\omega v}- \int_{-\infty}^{0}\! dv\, \alpha(v)  e^{J(v)}\,e^{-i\omega v}
\end{equation}
and we make use of of the standard result of the DCB theory\cite{Devoret_1990,Girvin_1990,Grabert_1991,Ingold_1992} for the current $I_{dc}$ in the presence of a dc voltage $V_{dc}$ only
\begin{equation}\label{Idc1}
I_{dc}(V_{dc})  =  \frac{e}{\hbar^2} 
\int_{-\infty}^{\infty} ds \,\alpha(s)\, e^{J(s)}
\left(e^{\frac{i}{\hbar}eV_{dc}s}-\hbox{c.c.}\right) .
\end{equation}
We also employ the Kramers-Kronig transformed current\cite{Grabert_2015}
\begin{equation}\label{IKK1}
I_{KK}(V_{dc}) = \frac{ie}{\hbar^2} 
\int_{-\infty}^{\infty}\!\!\! ds \,\operatorname{sign}(s)\,\alpha(s)\, e^{J(s)}
\left(e^{\frac{i}{\hbar}eV_{dc}s}+\hbox{c.c.}\right)
\end{equation}
which can be expressed in terms of $I_{dc}(V)$ by\cite{Tucker_1985,Grabert_2015}
\begin{equation}\label{IKK2}
I_{KK}(V)= P\int_{-\infty}^{\infty}\frac{dU}{\pi} \frac{I_{dc}(U)-G_TU}{U-V} .
\end{equation}
The relations (\ref{Idc1}) and (\ref{IKK1}) imply
\begin{eqnarray}\label{Idc2}\nonumber
&&\frac{\hbar^2}{e} I_{dc}\left(\frac{\hbar\omega}{e}\right)  =  
\int_{-\infty}^{\infty} ds \,\alpha(s)\, e^{J(s)}
\left(e^{i\omega s}-\hbox{c.c.}\right)\qquad\\ 
&&\qquad  = \int_{-\infty}^{0} ds \,\alpha(s)\, e^{J(s)}
\left(e^{i\omega s}-\hbox{c.c.}\right) \\ \nonumber
&&\qquad+  \int_{0}^{\infty} ds \,\alpha(s)\, e^{J(s)}
\left(e^{i\omega s}-\hbox{c.c.}\right)
\end{eqnarray}
and
\begin{eqnarray}\label{IKK3}\nonumber
&&i\frac{\hbar^2}{e} I_{KK}\!\left(\frac{\hbar\omega}{e}\right) = 
-\int_{-\infty}^{\infty}\!\!\! ds \,\operatorname{sign}(s)\alpha(s) e^{J(s)}\!
\left(e^{i\omega s}+\hbox{c.c.}\right) \\ 
&&\qquad = \int_{-\infty}^{0}\!\!\! ds \,\alpha(s)\, e^{J(s)}
\left(e^{i\omega s}+\hbox{c.c.}\right) \\ \nonumber
&&\qquad -\int_{0}^{\infty}\!\!\! ds \,\alpha(s)\, e^{J(s)}
\left(e^{i\omega s}+\hbox{c.c.}\right)
\end{eqnarray}
which can be combined to give
\begin{eqnarray}
&&\frac{\hbar^2}{2e}\left[ I_{dc}\left(\frac{\hbar\omega}{e}\right)\pm i I_{KK}\left(\frac{\hbar\omega}{e}\right) \right]\\ \nonumber
&& = \pm\int_{-\infty}^{0} ds \,\alpha(s)\, e^{J(s)}\,
e^{\pm i\omega s}\mp \int_{0}^{\infty} ds \,\alpha(s)\, e^{J(s)}\, e^{\mp i\omega s} .
\end{eqnarray}
This result can now we used to write the integral (\ref{intv2}) in the form
\begin{equation}\label{intv3}
L(\omega) =  \frac{\hbar^2}{2e}\left[ I_{dc}\left(\frac{\hbar\omega}{e}\right)- i I_{KK}\left(\frac{\hbar\omega}{e}\right) \right] .
\end{equation}

%%%%%%%%%%%%%%%%%%%%
\subsubsection{Evaluation of $S_{NT}(\omega)$ continued}
Using the results (\ref{HK3}) and (\ref{intv3}) we obtain from Eq.~(\ref{SNTomega2})
\begin{eqnarray}\label{SNTomega3}\nonumber
&&S_{NT}(\omega) =-\frac{\hbar^2C^2}{2e^3}\left(1-e^{-\beta\hbar\omega}\right) S_K(\omega)^2   \sum_{k=-\infty}^{\infty}J_k(a)^2  \\ \nonumber
&&\quad  \times
\bigg[ I_{dc}\!\left(V_{dc}+\frac{\hbar(k\Omega+\omega)}{e}\right) 
-I_{dc}\!\left(V_{dc}+\frac{\hbar(k\Omega-\omega)}{e}\right) \bigg]
\\ \nonumber
&&\quad +  \frac{\hbar C}{e} S_K(\omega)\,\left[1 -\omega\,C\, Z_t^{\prime\prime}(\omega)\right]\sum_{k=-\infty}^{\infty}J_k(a)^2 \\ \nonumber
&&\quad\times \bigg[ I_{KK}\!\left(\!V_{dc}+\frac{\hbar(k\Omega+\omega)}{e}\!\right)
\!+I_{KK}\!\left(\!V_{dc}+\frac{\hbar(k\Omega-\omega)}{e}\!\right) \\ 
&&\qquad
 -2I_{KK}\!\left(V_{dc}+\frac{k\hbar\Omega}{e}\right) 
 \bigg] .
\end{eqnarray}
Furthermore, by inserting the explicit expression  (\ref{SKofom}) for $S_K(\omega)$, we obtain for the frequency-dependent factors on the right-hand side of Eq.~(\ref{SNTomega3})
\begin{eqnarray}\nonumber
&&\frac{\hbar^2C^2}{2e^3}\left(1-e^{-\beta\hbar\omega}\right) S_K(\omega)^2 =
 \frac{2e[\omega C  Z_t^{\prime}(\omega)]^2 }{1- e^{-\beta\hbar\omega}}\\ 
&&\qquad = \frac{2e }{1- e^{-\beta\hbar\omega}}\, \Xi(\omega)^2\sin^2[\eta(\omega)]\end{eqnarray}
and
\begin{eqnarray}\nonumber
&&\frac{\hbar C}{e} S_K(\omega)\left[1 -\omega C  Z_t^{\prime\prime}(\omega)\right]=
  \frac{2e \omega C  Z_t^{\prime}(\omega) \left[1 -\omega C Z_t^{\prime\prime}(\omega)\right]}{1- e^{-\beta\hbar\omega}}\\ 
  && =  \frac{2e }{1- e^{-\beta\hbar\omega}}\, \Xi(\omega)^2 \sin[\eta(\omega)] \cos[\eta(\omega)]
\end{eqnarray}
where we have made use of the polar decomposition (\ref{polarde}).
This allows us to write the result (\ref{SNTomega3}) in the form
\begin{eqnarray}\label{SNTomega4}\nonumber
&&S_{NT}(\omega) =\frac{2e}{1- e^{-\beta\hbar\omega}}\,\Xi(\omega)^2 \sum_{k=-\infty}^{\infty}J_k(a)^2 \bigg\{ \sin^2[\eta(\omega)] \\ \nonumber
&&  \times
\bigg[ I_{dc}\!\left(\! V_{dc}+\frac{\hbar(k\Omega-\omega)}{e}\right) 
-I_{dc}\!\left(\! V_{dc}+\frac{\hbar(k\Omega+\omega)}{e}\right) \bigg]
\\ 
&& +  \frac{\sin[2\eta(\omega)]}{2}
\bigg[ I_{KK}\!\left(\!V_{dc}+\frac{\hbar(k\Omega-\omega)}{e}\!\right) \\  \nonumber
&&\quad
+I_{KK}\!\left(\!V_{dc}+\frac{\hbar(k\Omega+\omega)}{e}\!\right) 
 -2I_{KK}\!\left(V_{dc}+\frac{k\hbar\Omega}{e}\right) 
 \bigg]\bigg\} .
\end{eqnarray}
Clearly, also this contribution to the spectral function can be written in terms of the current-voltage relation $I_{dc}(V_{dc})$ in the absence of an ac driving voltage. 

One can show\cite{Frey_2016} that the contribution $S_{NT}(\omega)$ to the current noise spectrum comes from a cross-correlation between fluctuations of the tunneling current and the Johnson-Nyquist noise of the electromagnetic environment. The latter causes fluctuations of the voltage across the tunnel junction which then induce fluctuations of the tunneling current. Correspondingly, the result (\ref{SNTomega4}) combines frequency-dependent factors stemming from circuit theory and the finite-frequency junction admittance. 

In the absence of an ac voltage, i.e., when $a=0$, the expression (\ref{SNTomega4}) simplifies to read as
\begin{eqnarray}\label{SNTdc} \nonumber
&&S_{NT}(V_{dc},0,\omega)=  \frac{2e}{1- e^{-\beta\hbar\omega}}\,\Xi(\omega)^2\bigg\{\sin^2[\eta(\omega)]   \\ \nonumber
&&\quad\times\bigg[  I_{dc}\left(V_{dc}-\hbar\omega/e\right) 
-I_{dc}\left(V_{dc}+\hbar\omega/e\right) \bigg]
\\
&&\quad +  \frac{\sin[2\eta(\omega)]}{2}  \bigg[ I_{KK}\left(V_{dc}-\hbar\omega/e\right) \\ \nonumber
&&\quad 
+I_{KK}\left(V_{dc}+\hbar\omega/e\right)  -2I_{KK}\left(V_{dc}\right)  \bigg]\bigg\}
\end{eqnarray}
where we have again made the voltage dependence of the spectral function explicit. Clearly, the noise component $S_{NT}(\omega)$ is also present for tunnel junctions solely under dc bias.\cite{Lee_1996}  

Finally, when we compare the results (\ref{SNTomega4}) and (\ref{SNTdc}), we see that the contribution $S_{NT}(V_{dc},V_{ac},\omega)$ to the spectral function $S(V_{dc},V_{ac},\omega)$ is related to the corresponding contribution $S_{NT}(V_{dc},0,\omega)$ measured at dc bias by a photo-assisted tunneling relation
\begin{equation}\label{paNT}
S_{NT}(V_{dc},V_{ac},\omega)=\sum_{k=-\infty}^{\infty}J_k(a)^2\, S_{NT}(V_{dc}+k\hbar\Omega/e,0,\omega) .
\end{equation}

%%%%%%%%%%%%%%%%%
\subsection{Photo-assisted tunneling relation for \bm{$S(\omega)$}}
We have shown that the components $S_T(V_{dc},V_{ac},\omega)$ and $S_{NT}(V_{dc},V_{ac},\omega)$ of the spectral function $S(V_{dc},V_{ac},\omega)$ can be related by photo-assisted tunneling relations [given in Eqs.~(\ref{paT}) and (\ref{paNT})] to the corresponding components $S_T(V_{dc},0,\omega)$ and $S_{NT}(V_{dc},0,\omega)$ in the absence of an ac voltage $V_{ac}$.
The remaining component $S_N(V_{dc},V_{ac},\omega)$ determined in Eq.~(\ref{SNomega2}) is in fact independent of $V_{dc}$ and $V_{ac}$.
Nevertheless, we may formally write\begin{equation}\label{paN}
S_N(V_{dc},V_{ac},\omega)=\sum_{k=-\infty}^{\infty}J_k(a)^2\, S_N(V_{dc}+k\hbar\Omega/e,0,\omega)
\end{equation}
since\cite{Abramowitz}
\begin{equation}
\sum_{k=-\infty}^{\infty}J_k(a)^2 =1 .
\end{equation}
Therefore, also the total spectral density  $S(V_{dc},V_{ac},\omega)$ can be expressed through the spectrum in the absence of an ac voltage by a relation
\begin{equation}\label{paTR}
S(V_{dc},V_{ac},\omega)=\sum_{k=-\infty}^{\infty}J_k(a)^2\, S(V_{dc}+k\hbar\Omega/e,0,\omega)
\end{equation}
of the Tien-Gordon type. 
Hence, the spectral function $S(V_{dc},V_{ac},\omega)$ of an ac driven device is obtained as copies of the same quantity measured under dc bias, with dc voltages translated by shifts $n\hbar\Omega/e$ corresponding to the harmonics and weighted by Bessel functions.
This property has been noticed previously\cite{Parlavecchio_2015} for the spectrum of the tunneling current and it remains true for the spectrum of the measurable environmental current.

%%%%%%%%%%%%%%%%%%%%%%
%%%%%%%%%%%%%%%%%%%%%%
%%%%%%%%%%%%%%%%%%%%%%
\section{Specific electromagnetic environments}\label{sec:five}
In this section we specify the results derived in the previous sections for two experimentally relevant models of the electromagnetic environment.

%%%%%%%%%%%%%%%%%%%%%%
\subsection{Ohmic environment}

We first consider the special case of a purely Ohmic environment with an impedance $Z(\omega)=R$, which implies
\begin{equation}\label{ZtR}
Z_t(\omega)=\frac{R}{1-i\omega RC} .
\end{equation}
This gives
\begin{equation}
\frac{Z_t(\omega)}{Z(\omega)} = \frac{1}{1-i\omega RC}
\end{equation}
which has the modulus
\begin{equation}
\Xi(\omega)=\sqrt{\frac{1}{1+(\omega RC)^2}}
\end{equation}
and the phase
\begin{equation}
\arctan[\eta(\omega)]=\omega RC .
\end{equation}
Accordingly,
\begin{equation}
\sin[\eta(\omega)]= \frac{\omega RC}{\sqrt{1+(\omega RC)^2}}=\omega RC\,\Xi(\omega)  
\end{equation}
and
\begin{equation}\label{cosR}
\cos[\eta(\omega)]=\frac{1}{\sqrt{1+(\omega RC)^2}}=\Xi(\omega) .
\end{equation}
We then obtain from Eq.~(\ref{SNomega2}) for the Johnson-Nyquist part of the spectral function
\begin{equation}\label{SNR}
S_N(\omega)=  \frac{2\hbar \omega^3 C^2 R}{\left(1- e^{-\beta\hbar\omega}\right)\left[1+(\omega RC)^2  \right]} 
\end{equation} 
which is the  Johnson-Nyquist current noise 
\begin{equation}
S_{I_NI_N}(\omega)=\frac{2\hbar\omega}{1-\exp(-\beta\hbar\omega)}\frac{1}{R}
\end{equation}
of the resistor $R$ scaled by the factor $(\omega RC)^2/[1+(\omega RC)^2]=[\Xi(\omega)\omega RC\,]^2$.

The shot noise part of the spectral function $S(\omega)$ is obtained from Eqs.~(\ref{STomega5}) and (\ref{ZtR})
in the form
\begin{eqnarray}\label{STR}
&&S_T(\omega)  = \frac{e}{1+(\omega RC)^2 }
\sum_{k=-\infty}^{\infty}J_k(a)^2
\\  \nonumber
&& \times 
\bigg\{\frac{I_{dc}\left(V_{dc}+\hbar(k\Omega+\omega)/e\right)}{1-e^{-\beta\left[eV_{dc}+\hbar(k\Omega+\omega)\right]}}
+ \frac{I_{dc}\left(V_{dc}+\hbar(k\Omega-\omega)/e\right)}{e^{\beta\left[eV_{dc}+\hbar(k\Omega-\omega)\right]}-1}  \bigg\}
\end{eqnarray}
which corresponds to the shot noise of the tunneling current scaled by the factor $1/[1+(\omega RC)^2]$=$\Xi(\omega)^2$.

Finally, the cross-correlation between the noise from the resistor $R$ and the shot noise of the junction leads to the contribution
\begin{eqnarray}\label{SNTR}\nonumber
&&S_{NT}(\omega) =\frac{2e}{(1- e^{-\beta\hbar\omega})}\frac{ \omega RC }{[1+(\omega RC)^2]^2}  \sum_{k=-\infty}^{\infty}J_k(a)^2  \\ 
&&\quad  \times \bigg\{\omega RC
\bigg[ I_{dc}\!\left(V_{dc}+\frac{\hbar(k\Omega-\omega)}{e}\right) \\ \nonumber
&&\quad 
-I_{dc}\!\left(V_{dc}+\frac{\hbar(k\Omega+\omega)}{e}\right) \bigg]
\! +I_{KK}\!\left(\!V_{dc}+\frac{\hbar(k\Omega-\omega)}{e}\!\right)
  \\ \nonumber
&&\quad  
+I_{KK}\!\left(\!V_{dc}+\frac{\hbar(k\Omega+\omega)}{e}\!\right)
 -2I_{KK}\!\left(V_{dc}+\frac{k\hbar\Omega}{e}\right) 
 \bigg\} .
\end{eqnarray}
To derive this result, we have combined Eq.~(\ref{SNTomega4})  and Eqs.~(\ref{ZtR}) -- (\ref{cosR}).

Let us now briefly address the low impedance limit of the Ohmic model.
Frequently, environmental impedances are quite small, in the range of 50 $\Omega$, which is the order of magnitude of typical transmission line impedances.
In the limit $R\ll R_K$ of a low impedance environment   the contributions $S_N(\omega)$ and $S_{NT}(\omega)$ to the spectral function become negligible and we simply have
\begin{eqnarray}
&&S(\omega)  = e
\sum_{k=-\infty}^{\infty}J_k(a)^2
\\  \nonumber
&& \times 
\bigg\{\frac{I_{dc}\left(V_{dc}+\hbar(k\Omega+\omega)/e\right)}{1-e^{-\beta\left[eV_{dc}+\hbar(k\Omega+\omega)\right]}}
+ \frac{I_{dc}\left(V_{dc}+\hbar(k\Omega-\omega)/e\right)}{e^{\beta\left[eV_{dc}+\hbar(k\Omega-\omega)\right]}-1}  \bigg\}
\end{eqnarray}
which is the shot noise of an ac driven tunnel junction in the absence of effects of the electromagnetic environment.\cite{Tucker_1985,Lesovik_1994} This is natural since in this limit the influence of the electromagnetic environment vanishes. There is no sizable DCB effect,  and the dc current-voltage characteristic of the tunnel junction is approximately linear\cite{Ingold_1992}, i.e., $I_{dc}(V_{dc})= G_T V_{dc}$.

%%%%%%%%%%%%%%%%
\subsection{Tunnel junction driven through a resonator }
We consider now a tunnel junction driven through an $LC$-resonator as studied recently.\cite{Parlavecchio_2015,Altimiras_2014, Grabert_2015} 
The environmental impedance is assumed to be of the form
\begin{equation}
Z(\omega)=R-i\omega L
\end{equation}
with an Ohmic lead resistance $R$ and an inductance $L$. The resonance frequency of the $LC$-resonator is 
\begin{equation}
\omega_0=\frac{1}{\sqrt{LC}}
\end{equation}
and it has the characteristic impedance 
\begin{equation}
Z_c=\sqrt{\frac{L}{C}}
\end{equation}
implying a quality factor $Q_f=Z_c/R$. We shall also use the loss factor
\begin{equation}
\gamma=\frac{1}{Q_f}=\frac{R}{Z_c} .
\end{equation}
For this circuit the ratio between the total impedance (\ref{Zt}) and the impedance $Z(\omega)$ of the leads is given by
\begin{equation}
\frac{Z_t(\omega)}{Z(\omega)}=\frac{1}{1-i\omega RC -\omega^2 LC}=\frac{\omega_0^2}{\omega_0^2-\omega^2-i\gamma\omega_0\omega}
\end{equation}
which implies a modulus
\begin{equation}\label{Xires}
\Xi(\omega)=\frac{\omega_0^2}{\sqrt{\left(\omega_0^2-\omega^2\right)^2+\left(\gamma\omega_0\omega\right)^2}} 
\end{equation}
and a phase
\begin{equation} \label{etares}
\eta(\omega)=\arctan\left(\frac{\gamma\omega_0\omega}{\omega_0^2-\omega^2}\right)
\end{equation}
where the values of $\arctan$ are to be chosen in the interval $[0,\pi)$. We then have
\begin{equation}
\sin[\eta(\omega)]=\frac{\gamma\omega_0\omega}{\sqrt{(\omega_0^2-\omega^2)^2+(\gamma\omega_0\omega)^2}}
\end{equation}
and
\begin{equation}
\cos[\eta(\omega)]=\frac{\omega_0^2-\omega^2}{\sqrt{(\omega_0^2-\omega^2)^2+(\gamma\omega_0\omega)^2}} .
\end{equation}
The effects of the electromagnetic environment are most pronounced when the frequency $\omega$ coincides with the resonance frequency $\omega_0$ of the $LC$-resonator.   We obtain from Eqs.~(\ref{Xires}) and (\ref{etares}) for $\omega=\omega_0$
\begin{equation}\label{res}
\Xi(\omega_0) = Q_f, \quad \eta(\omega_0)=\frac{\pi}{2} .
\end{equation}
For this particular frequency we find from Eq.~(\ref{SNomega3})
\begin{equation}\label{SNLC}
S_N(\omega_0)= \frac{2\hbar\omega_0 }{1- e^{-\beta\hbar\omega_0}}   \frac{1}{R}
\end{equation}
where we have made use of $\Xi(\omega_0)\omega_0 R  C=1$. This corresponds to the Johnson-Nyquist noise of the lead resistance $R$.

The shot noise contribution (\ref{STomega5}) takes the form
\begin{eqnarray}\label{STLC} \nonumber
S_T(\omega_0) & =&  eQ_f^2
\sum_{k=-\infty}^{\infty}\! J_k(a)^2
\bigg\{\frac{I_{dc}\left(V_{dc}+\hbar(k\Omega+\omega_0)/e\right)}{1-e^{-\beta\left[eV_{dc}+\hbar(k\Omega+\omega_0)\right]}}\\ 
&& \qquad
+ \frac{I_{dc}\left(V_{dc}+\hbar(k\Omega-\omega_0)/e\right)}{e^{\beta\left[eV_{dc}+\hbar(k\Omega-\omega_0)\right]}-1}  \bigg\}
\end{eqnarray}
and is thus strongly enhanced by the factor $Q_f^2$ relative to the noise of the tunneling current through the junction.

Finally, the cross-correlation between the shot noise and the resistor noise gives rise to the contribution
\begin{eqnarray}\label{SNTLC}
&&S_{NT}(\omega_0) =\frac{2e}{1- e^{-\beta\hbar\omega_0}}\,Q_f^2 \sum_{k=-\infty}^{\infty}J_k(a)^2  \\ \nonumber
&& \times
\bigg[ I_{dc}\!\left(\! V_{dc}+\frac{\hbar(k\Omega-\omega_0)}{e}\right) 
-I_{dc}\!\left(\! V_{dc}+\frac{\hbar(k\Omega+\omega_0)}{e}\right)  \bigg]
\end{eqnarray} 
where we have particularized the result (\ref{SNTomega4}) using the specific values for $\omega=\omega_0$ given in Eq.~(\ref{res}).

Finally, let us  consider the zero temperature spectral function $S(V_{dc},0,\omega_0)$ in the absence of an ac voltage. From Eqs.~(\ref{SNLC}) -- (\ref{SNTLC}) we obtain for $T=0$ and $V_{ac}=0$
\begin{equation}
S_N(\omega_0)=\frac{2\hbar\omega_0 }{R} ,
\end{equation}
\begin{eqnarray}\nonumber
&&S_T(V_{dc},0,\omega_0)=  e Q_f^2\bigg[\Theta(eV_{dc}+\hbar\omega_0)I_{dc}(V_{dc}+\hbar\omega_0/e)  \\ 
&&\qquad +\Theta(\hbar\omega_0-eV_{dc})I_{dc}(\hbar\omega_0/e-V_{dc})\bigg]
\end{eqnarray}
and
\begin{eqnarray}
&&S_{NT}(V_{dc},0,\omega_0)\\ \nonumber
&&\quad=  2e Q_f^2\bigg[I_{dc}(V_{dc}-\hbar\omega_0/e) -   I_{dc}(V_{dc}+\hbar\omega_0/e)\bigg]
\end{eqnarray}
where $\Theta(x)$ is the unit step function. This shows that the contribution $S_{NT}(\omega)$ can be of the same order of magnitude as $S_T(\omega)$ already for junctions under dc bias. The corresponding results for finite ac voltage $V_{ac}$ can readily be obtained by means of the photo-assisted tunneling relation (\ref{paTR}).

%%%%%%%%%%%%%%%%%%%
%%%%%%%%%%%%%%%%%%%
%%%%%%%%%%%%%%%%%%%
\section{Conclusions}\label{sec:six}
We have determined the spectrum of current fluctuations of a tunnel junction biased by an external voltage applied via an environmental impedance. Apart from the noise components arising from the Johnson-Nyquist noise of the electromagnetic environment and the shot noise of the tunneling element, respectively, there is a third component which comes from the correlation between fluctuations of the tunneling current and the environmental Johnson-Nyquist noise. This component can also be interpreted as a tunneling modification of the current fluctuations seen in the outer circuit which originate from the current fluctuations generated by the environmental impedance. Tunneling opens up a second channel for the transmission of these fluctuations to the external leads apart from the transfer via the junction capacitance.

Regarding the noise of ac driven tunnel junctions we have demonstrated that the spectral function for finite ac voltage can be obtained from the spectral function at dc bias by means of a photo-assisted tunneling relation of the Tien-Gordon type. This relation is formally valid for each of the three noise components separately. Studying specific models of the electromagnetic environment we have shown that the third noise component can be as significant as the two other components. Experimental consequences of this fact remain to be analyzed. Also, the results obtained here for the current fluctuations can be combined with previous findings concerning the average current to examine the validity of fluctuation dissipation relations and related questions. This and other aspects of the theory will be addressed elsewhere.

%%%%%%%%%%%%%%%%%%%%%%%
%%%%%%%%%%%%%%%%%%%%%%%
%%%%%%%%%%%%%%%%%%%%%%%
\appendix
\begin{widetext}
\section{Evaluation of the current autocorrelation function \bm{$C(t+s,t)$}}\label{app:A}
Employing the expansion (\ref{IenvHeis2}) of the current operator $I_{env}(t)$ up to second order in the tunneling Hamiltonian $H_T$, the corresponding expansion of the current autocorrelation function $C(t+s,t)=\langle I_{env}(t+s)I_{env}(t)\rangle-\langle I_{env}(t+s)\rangle \langle I_{env}(t)\rangle $  is readily obtained in the form
\begin{eqnarray}\label{Ienvcorr}
&&C(t+s,t)=  \Big\langle\Big\{ \check I_{env}(t+s)   +\frac{i}{\hbar} \int_{0}^{\infty} du\,   \left[\check\Theta(t+s-u)\, e^{-i\check\varphi(t+s-u)}+ \hbox{H.c.}\, ,  \check I_{env}  (t+s)\right] \\ \nonumber
&&\quad -\frac{1}{\hbar^2}\int_{0}^{\infty} du \int_{0}^{\infty} dv\,   \Big[\check\Theta(t+s-u-v)\, e^{-i\check\varphi(t+s-u-v)}+ \hbox{H.c.}\, , \Big[\check\Theta(t+s-u)\, e^{-i\check\varphi(t+s-u)}+ \hbox{H.c.}\, ,  \check I_{env}  (t+s)\Big] \Big]\Big\}
\\ \nonumber
&&\quad\times\Big\{\check I_{env}(t)  +\frac{i}{\hbar} \int_{0}^{\infty} du\,   \left[\check\Theta(t-u)\, e^{-i\check\varphi(t-u)}+ \hbox{H.c.}\, ,  \check I_{env}  (t)\right] \\ \nonumber
&&\quad -\frac{1}{\hbar^2}\int_{0}^{\infty} du \int_{0}^{\infty} dv\,   \Big[\check\Theta(t-u-v)\, e^{-i\check\varphi(t-u-v)}+ \hbox{H.c.}\, ,  \Big[\check\Theta(t-u)\, e^{-i\check\varphi(t-u)}+ \hbox{H.c.}\, ,  \check I_{env}  (t)\Big] \Big]\Big\}\Big\rangle\\ \nonumber
%%%%%%%%%%%%%%%%
&&\quad-\Big\langle \check I_{env}(t+s)   +\frac{i}{\hbar} \int_{0}^{\infty} du\,   \left[\check\Theta(t+s-u)\, e^{-i\check\varphi(t+s-u)}+ \hbox{H.c.}\, ,  \check I_{env}  (t+s)\right] \\ \nonumber
&&\quad -\frac{1}{\hbar^2}\int_{0}^{\infty} du \int_{0}^{\infty} dv\,   \Big[\check\Theta(t+s-u-v)\, e^{-i\check\varphi(t+s-u-v)} + \hbox{H.c.}\, , \Big[\check\Theta(t+s-u)\, e^{-i\check\varphi(t+s-u)}+ \hbox{H.c.}\, ,  \check I_{env}  (t+s)\Big] \Big]\Big\rangle\\ \nonumber
&&\quad\times\Big\langle \check I_{env}(t)  +\frac{i}{\hbar} \int_{0}^{\infty} du\,   \left[\check\Theta(t-u)\, e^{-i\check\varphi(t-u)}+ \hbox{H.c.}\, ,  \check I_{env}  (t)\right] \\ \nonumber
&&\quad -\frac{1}{\hbar^2}\int_{0}^{\infty} du \int_{0}^{\infty} dv\,   \Big[\check\Theta(t-u-v)\, e^{-i\check\varphi(t-u-v)}+ \hbox{H.c.}\, ,  \Big[\check\Theta(t-u)\, e^{-i\check\varphi(t-u)}+ \hbox{H.c.}\, ,  \check I_{env}  (t)\Big] \Big]\Big\rangle .
\end{eqnarray}
\subsection{Evaluation of quasiparticle averages}
Since $H_{el}$ and $H_{env}$ decouple in the absence of tunneling, each term of Eq.~(\ref{Ienvcorr}) factorizes into an average over quasiparticle operators and an average over the electromagnetic environment. The arising quasiparticle averages coincide with those known from the standard $P(E)$ theory\cite{Devoret_1990,Girvin_1990,Grabert_1991,Ingold_1992} and can be expressed in terms of the correlator of tunneling operators $\alpha(t)$ introduced in Eq.~(\ref{alpha}). Since only terms correlating a tunneling operator $\Theta$ with its adjoint $\Theta^{\dagger}$ give a finite contribution we obtain
\begin{eqnarray}
&&C(t+s,t) = \Big\langle \tilde I_{env}(t+s)\tilde I_{env}(t) \Big\rangle 
  -\frac{1}{\hbar^2}\int_{0}^{\infty} du \int_{0}^{\infty} dv\,\\ \nonumber
&&\qquad \Big\{ \alpha(-v)\Big(\Big\langle \tilde I_{env}(t+s)  \, e^{-i\check\varphi(t-u-v)}\, \Big[ e^{i\check\varphi(t-u)} ,  \tilde I_{env}  (t) \Big]\Big\rangle 
+ \Big\langle \tilde I_{env}(t+s)      e^{i\check\varphi(t-u-v)} \Big[ e^{-i\check\varphi(t-u)} ,  \tilde I_{env}  (t) \Big]\Big\rangle \Big)\\ \nonumber
&&\qquad-\alpha(v) \Big(\Big\langle \tilde I_{env}(t+s)     \Big[ e^{i\check\varphi(t-u)} ,  \tilde I_{env}  (t)\Big] e^{-i\check\varphi(t-u-v)}\Big\rangle +\Big\langle\tilde I_{env}(t+s)     \Big[ e^{-i\check\varphi(t-u)} ,   \tilde I_{env}  (t)\Big] e^{i\check\varphi(t-u-v)}\Big\rangle\Big)\Big\} \\ \nonumber
&&\quad  -\frac{1}{\hbar^2}\int_{0}^{\infty} du \int_{0}^{\infty} dv\, \\ \nonumber
&& \qquad\Big\{ \alpha(-v)\Big(\Big\langle e^{-i\check\varphi(t+s-u-v)}\, \Big[ e^{i\check\varphi(t+s-u)} ,  \tilde I_{env}  (t+s) \Big]  \tilde I_{env}(t)  \Big\rangle 
+ \Big\langle   e^{i\check\varphi(t+s-u-v)} \Big[ e^{-i\check\varphi(t+s-u)} ,  \tilde I_{env}  (t+s) \Big]\tilde I_{env}(t)    \Big\rangle \Big)\\ \nonumber
&&\qquad-\alpha(v) \Big(\Big\langle   \Big[ e^{i\check\varphi(t+s-u)} ,   \tilde I_{env}  (t+s)\Big] e^{-i\check\varphi(t+s-u-v)}\tilde I_{env}(t)   \Big\rangle 
+\Big\langle    \Big[ e^{-i\check\varphi(t+s-u)} ,  \tilde I_{env}  (t+s)\Big] e^{i\check\varphi(t+s-u-v)}   \tilde I_{env}(t) \Big\rangle\Big)\Big\} \\ \nonumber
&&\quad
-\frac{1}{\hbar^2} \int_{0}^{\infty} du\,    \int_{0}^{\infty} dv\,  \alpha(s-u+v)\\ \nonumber
&&\qquad\Big\{ \Big\langle  \left[ e^{-i\check\varphi(t+s-u)} ,  \tilde I_{env}  (t+s)\right] 
  \left[ e^{i\check\varphi(t-v)} ,  \tilde I_{env}  (t)\right]  \Big\rangle 
+ \Big\langle  \left[ e^{i\check\varphi(t+s-u)} ,  \tilde I_{env}  (t+s)\right] 
  \left[ e^{-i\check\varphi(t-v)} ,  \tilde I_{env}  (t)\right]  \Big\rangle \Big\}
\end{eqnarray}
where we have also inserted the decomposition (\ref{Ienvcheck}) of $\check I_{env}(t)$.

To proceed, we now insert the decomposition (\ref{phicheck2}) of $\check\varphi(t)$ and the representation (\ref{tildeIenv}) of $\tilde I_{env}(t)$ which gives 
\begin{eqnarray}\label{Ienvcorr3}
&&C(t+s,t)=  \frac{\hbar^2C^2}{e^2}\Big\langle \ddot{ \tilde\varphi}(t+s)\ddot{ \tilde\varphi}(t) \Big\rangle 
 -\frac{C^2}{e^2}\int_{0}^{\infty} du \int_{0}^{\infty} dv\,\\ \nonumber
&& \qquad \Big\{ \alpha(-v)\Big(\Big\langle \ddot{ \tilde\varphi}(t+s)  \, e^{-i\tilde\varphi(t-u-v)}\, \Big[ e^{i\tilde\varphi(t-u)} ,  \ddot{ \tilde\varphi}  (t) \Big]\Big\rangle   e^{-i\bar\varphi(t-u-v)}e^{i\bar\varphi(t-u)}
\\ \nonumber
&& \qquad\quad
+ \Big\langle \ddot{ \tilde\varphi}(t+s)      e^{i\tilde\varphi(t-u-v)} \Big[ e^{-i\tilde\varphi(t-u)} ,  \ddot{ \tilde\varphi}  (t) \Big]\Big\rangle  e^{i\bar\varphi(t-u-v)} e^{-i\bar\varphi(t-u)}
\Big)
\\ \nonumber
&& \qquad
-\alpha(v) \Big(\Big\langle \ddot{ \tilde\varphi}(t+s)     \Big[ e^{i\tilde\varphi(t-u)} ,  \ddot{ \tilde\varphi}  (t)\Big] e^{-i\tilde\varphi(t-u-v)}\Big\rangle e^{i\bar\varphi(t-u)} e^{-i\bar\varphi(t-u-v)}
\\ \nonumber
&& \qquad\quad
+\Big\langle\ddot{ \tilde\varphi}(t+s)     \Big[ e^{-i\tilde\varphi(t-u)} ,   \ddot{ \tilde\varphi}  (t)\Big] e^{i\tilde\varphi(t-u-v)}\Big\rangle  e^{-i\bar\varphi(t-u)} e^{i\bar\varphi(t-u-v)}
\Big)\Big\} 
\\ \nonumber
&& \quad -\frac{C^2}{e^2}\int_{0}^{\infty} du \int_{0}^{\infty} dv\, \\ \nonumber
&& \qquad\Big\{ \alpha(-v)\Big(\Big\langle e^{-i\tilde\varphi(t+s-u-v)}\, \Big[ e^{i\tilde\varphi(t+s-u)} ,  \ddot{ \tilde\varphi}  (t+s) \Big]  \ddot{ \tilde\varphi}(t)  \Big\rangle   e^{-i\bar\varphi(t+s-u-v)} e^{i\bar\varphi(t+s-u)}
\\ \nonumber
&&\qquad\quad
+ \Big\langle   e^{i\tilde\varphi(t+s-u-v)} \Big[ e^{-i\tilde\varphi(t+s-u)} ,  \ddot{ \tilde\varphi}  (t+s) \Big]\ddot{ \tilde\varphi}(t)    \Big\rangle   e^{i\bar\varphi(t+s-u-v)} e^{-i\bar\varphi(t+s-u)}
\Big)
\\ \nonumber
&&\qquad-\alpha(v) \Big(\Big\langle   \Big[ e^{i\tilde\varphi(t+s-u)} ,   \ddot{ \tilde\varphi}  (t+s)\Big] e^{-i\tilde\varphi(t+s-u-v)}\ddot{ \tilde\varphi}(t)   \Big\rangle  e^{i\bar\varphi(t+s-u)} e^{-i\bar\varphi(t+s-u-v)}
\\ \nonumber
&&\qquad\quad
+\Big\langle    \Big[ e^{-i\tilde\varphi(t+s-u)} ,  \ddot{ \tilde\varphi}  (t+s)\Big] e^{i\tilde\varphi(t+s-u-v)}   \ddot{ \tilde\varphi}(t) \Big\rangle e^{-i\bar\varphi(t+s-u)} e^{i\bar\varphi(t+s-u-v)} 
\Big)\Big\} \\ \nonumber
&&\quad
-\frac{C^2}{e^2} \int_{0}^{\infty} du\,    \int_{0}^{\infty} dv\,  \alpha(s-u+v)\\ \nonumber
&&\qquad\Big\{ \Big\langle  \left[ e^{-i\tilde\varphi(t+s-u)} ,  \ddot{ \tilde\varphi}  (t+s)\right] 
  \left[ e^{i\tilde\varphi(t-v)} ,  \ddot{ \tilde\varphi}  (t)\right]  \Big\rangle  e^{-i\bar\varphi(t+s-u)}  e^{i\bar\varphi(t-v)}
\\ \nonumber
&&\qquad\quad
+ \Big\langle  \left[ e^{i\tilde\varphi(t+s-u)} ,  \ddot{ \tilde\varphi}  (t+s)\right] 
  \left[ e^{-i\tilde\varphi(t-v)} ,  \ddot{ \tilde\varphi}  (t)\right]  \Big\rangle e^{i\bar\varphi(t+s-u)}  e^{-i\bar\varphi(t-v)}\Big\} .
\end{eqnarray}
Here, all averages over the electromagnetic environment are expressed in terms of the phase operator $\tilde\varphi(t)$ of the undriven circuit in the absence of tunneling.

\subsection{Evaluation of averages over the electromagnetic environment}

To evaluate the averages over the electromagnetic environment in Eq.~(\ref{Ienvcorr3}) we make use of the phase-phase correlation function (\ref{Joft}) and the Gaussian statistics of phase fluctuations in the absence of tunneling and driving. We first note that Eq.~(\ref{Joft}) implies
\begin{equation}\label{ddJoft}
\ddot J(t)=\left\langle   \ddot{\tilde\varphi}(t) \tilde\varphi(0)  \right\rangle
\end{equation}
and
\begin{equation}\label{ddddJoft}
\ddddot J(t)=\left\langle   \ddot{\tilde\varphi}(t) \ddot{\tilde\varphi}(0)  \right\rangle .
\end{equation}
Accordingly, the first term on the right-hand side of Eq.~(\ref{Ienvcorr3}) may be written as
\begin{equation}
\frac{\hbar^2}{e^2}C^2 \Big\langle\ddot{ \tilde\varphi}(t+s) \ddot{ \tilde\varphi}(t) \Big\rangle =
\frac{\hbar^2}{e^2}C^2 \ddddot J(s) .
\end{equation}
To evaluate the other terms, we have to calculate the averages
\begin{equation}\label{phaseav}
 \Big\langle \ddot{ \tilde\varphi}(t+s)  \, e^{-i\tilde\varphi(t-u-v)}\, \Big[ e^{i\tilde\varphi(t-u)} ,  \ddot{ \tilde\varphi}  (t) \Big]\Big\rangle 
=- 2i e^{J(-v)} \ddot J^{\prime\prime}(u)\left[\ddot J(s+u+v)-\ddot J(s+u)\right]
\end{equation} 
\begin{equation}
 \Big\langle \ddot{ \tilde\varphi}(t+s)  \,    e^{i\tilde\varphi(t-u-v)} \Big[ e^{-i\tilde\varphi(t-u)} ,  \ddot{ \tilde\varphi}  (t) \Big]\Big\rangle 
= -2i e^{J(-v)}\ddot J^{\prime\prime}(u)\left[\ddot J(s+u+v)-J(s+u)\right]
\end{equation} 
\begin{equation}
\Big\langle \ddot{ \tilde\varphi}(t+s)     \Big[ e^{i\tilde\varphi(t-u)} ,  \ddot{ \tilde\varphi}  (t)\Big] e^{-i\tilde\varphi(t-u-v)}\Big\rangle 
= 2ie^{J(v)}\ddot J^{\prime\prime}(u)\left[\ddot J(s+u)-\ddot J(s+u+v)\right]
\end{equation} 
\begin{equation} 
\Big\langle\ddot{ \tilde\varphi}(t+s)     \Big[ e^{-i\tilde\varphi(t-u)} ,   \ddot{ \tilde\varphi}  (t)\Big] e^{i\tilde\varphi(t-u-v)}\Big\rangle 
= 2ie^{J(v)}\ddot J^{\prime\prime}(u)\left[\ddot J(s+u)-\ddot J(s+u+v)\right]
\end{equation} 
\begin{equation} 
 \Big\langle e^{-i\tilde\varphi(t+s-u-v)}\, \Big[ e^{i\tilde\varphi(t+s-u)} ,  \ddot{ \tilde\varphi}  (t+s) \Big]  \ddot{ \tilde\varphi}(t)  \Big\rangle 
=-2i e^{J(-v)}\ddot J^{\prime\prime}(u)\left[\ddot J(s-u-v)-\ddot J(s-u)\right]
\end{equation} 
\begin{equation}
 \Big\langle   e^{i\tilde\varphi(t+s-u-v)} \Big[ e^{-i\tilde\varphi(t+s-u)} ,  \ddot{ \tilde\varphi}  (t+s) \Big]\ddot{ \tilde\varphi}(t)    \Big\rangle  
= -2i e^{J(-v)}\ddot J^{\prime\prime}(u)\left[\ddot J(s-u-v)-\ddot J(s-u)\right]
\end{equation} 
\begin{equation}
\Big\langle   \Big[ e^{i\tilde\varphi(t+s-u)} ,   \ddot{ \tilde\varphi}  (t+s)\Big] e^{-i\tilde\varphi(t+s-u-v)}\ddot{ \tilde\varphi}(t)   \Big\rangle 
= 2ie^{J(v)}\ddot J^{\prime\prime}(u)\left[\ddot J(s-u)-\ddot J(s-u-v) \right]
\end{equation} 
\begin{equation} 
\Big\langle    \Big[ e^{-i\tilde\varphi(t+s-u)} ,  \ddot{ \tilde\varphi}  (t+s)\Big] e^{i\tilde\varphi(t+s-u-v)}   \ddot{ \tilde\varphi}(t) \Big\rangle 
= 2ie^{J(v)}\ddot J^{\prime\prime}(u)\left[\ddot J(s-u)-\ddot J(s-u-v)\right]
\end{equation} 
\begin{equation}
\Big\langle  \left[ e^{-i\tilde\varphi(t+s-u)} ,  \ddot{ \tilde\varphi}  (t+s)\right] 
  \left[ e^{i\tilde\varphi(t-v)} ,  \ddot{ \tilde\varphi}  (t)\right]  \Big\rangle  
= -2i e^{J(s-u+v)}\ddot J^{\prime\prime}(u)\left[\ddot J(-v)-\ddot J(v)\right]
\end{equation} 
\begin{equation}\label{phaseavfin}
\Big\langle  \left[ e^{i\tilde\varphi(t+s-u)} ,  \ddot{ \tilde\varphi}  (t+s)\right] 
  \left[ e^{-i\tilde\varphi(t-v)} ,  \ddot{ \tilde\varphi}  (t)\right]  \Big\rangle 
= -2ie^{J(s-u+v)} \ddot J^{\prime\prime}(u)\left[\ddot J(-v)-\ddot J(v)\right]
\end{equation}
where we have used the generalized Wick theorem for Gaussian processes\cite{Louisell_1990} to obtain the right hand sides. 

To indicate how the evaluation proceeds, we give intermediate results for Eq. (\ref{phaseav}). The Gaussian statistics implies
\begin{equation}
\Big\langle \ddot{ \tilde\varphi}(t+s)  \, e^{-i\tilde\varphi(t-u-v)}\, \Big[ e^{i\tilde\varphi(t-u)} ,  \ddot{ \tilde\varphi}  (t) \Big]\Big\rangle = i\Big\langle  \Big[ \tilde\varphi(t-u) ,  \ddot{ \tilde\varphi}  (t) \Big]\Big\rangle \Big\langle \ddot{ \tilde\varphi}(t+s)  \, e^{-i\tilde\varphi(t-u-v)}\,  e^{i\tilde\varphi(t-u)} \Big\rangle  .
\end{equation}
The two averages on the right-hand side of this relation can be related to the phase-phase correlation function (\ref{Joft}) by means of
\begin{equation}
\Big\langle  \left[\tilde\varphi(t) ,  \ddot{ \tilde\varphi}  (s)\right] \Big\rangle 
=\ddot J(t-s) - \ddot J(s-t)   = 2i \ddot J^{\prime\prime}(t-s) 
\end{equation}
where $ J^{\prime\prime}(t)$ denotes the imaginary part of $J(t)$, and
\begin{equation}
\Big\langle \ddot{ \tilde\varphi}(t)  \, e^{-i\tilde\varphi(u)}\,  e^{i\tilde\varphi(v)} \Big\rangle =-ie^{J(u-v)}\left[ \ddot J(t-u)-\ddot J(t-v)\right] .
\end{equation}
These relations combine to give the result (\ref{phaseav}).

Inserting the findings (\ref{phaseav}) -- (\ref{phaseavfin}) into Eq.~(\ref{Ienvcorr3}) we obtain the current autocorrelation function in the form (\ref{Ienvcorr6}) given in the main text.

%%%%%%%%%%%%%%%%%%%%%%%
%%%%%%%%%%%%%%%%%%%%%%%
%%%%%%%%%%%%%%%%%%%%%%%
\section{Determination of the spectral function \bm{$S(\omega)$}}\label{app:B}
The Fourier representation (\ref{Koft}) of $K(s)=-\ddot J(s)$ implies
\begin{equation}
\ddot J(s)-\ddot J(s+v) =-\int_{-\infty}^{\infty}\! \frac{d \nu}{2\pi}\, S_K(\nu) e^{-i\nu s}\left[1-e^{-i\nu v}\right]
\end{equation}
and
\begin{equation}
2i\ddot J^{\prime\prime}(s) =\ddot J(s)-\ddot J(-s)=-\int_{-\infty}^{\infty}\, \frac{d \mu}{2\pi}\, \left(1-e^{-\beta\hbar\mu}\right) S_K(\mu)\,e^{-i\mu s}  .
\end{equation}
Inserting these representations into Eq.~(\ref{C0ofs})
we obtain
\begin{eqnarray}\label{C0ofs2}
&&C_0(s)  =   \frac{\hbar^2C^2}{e^2}\int_{-\infty}^{\infty}\frac{d\omega}{2\pi}\,\omega^2 S_K(\omega)\, e^{-i\omega s}
\\ \nonumber
&&+\frac{C^2}{e^2}\int_{0}^{\infty} du  \int_{-\infty}^{\infty}\, \frac{d \mu}{2\pi}\, \left(1-e^{-\beta\hbar\mu}\right) S_K(\mu)\,e^{-i\mu u} \int_{0}^{\infty}\! dv \left(\alpha(v) 
 e^{J(v)} -\hbox{c.c.}\right)\int_{-\infty}^{\infty} \frac{d \nu}{2\pi}\, S_K(\nu)\,  e^{-i\nu( s+u)}\\  \nonumber
&& \quad \times \left[1-e^{-i\nu v}\right] 
 \sum_{k=-\infty}^{\infty}J_k(a)^2
\left( e^{\frac{i}{\hbar}eV_{dc}v}e^{ik\Omega v} +\hbox{c.c.}\right)
\\ \nonumber
&&  +\frac{C^2}{e^2}\int_{0}^{\infty} du  \int_{-\infty}^{\infty}\, \frac{d \mu}{2\pi}\, \left(1-e^{-\beta\hbar\mu}\right) S_K(\mu)\,e^{-i\mu u}\int_{0}^{\infty}\! dv \left(\alpha(v)  e^{J(v)}-\hbox{c.c.}\right)\int_{-\infty}^{\infty} \frac{d \nu}{2\pi}\, S_K(\nu)\, e^{-i\nu( s-u)} \\ \nonumber
&& \quad\times
\left[1-e^{i\nu v}\right]  \sum_{k=-\infty}^{\infty}J_k(a)^2
\left( e^{\frac{i}{\hbar}eV_{dc}v}\, e^{ik\Omega v}+\hbox{c.c.}\right)
\\ \nonumber
&&
-\frac{C^2}{e^2} \int_{0}^{\infty} du 
\int_{-\infty}^{\infty} \frac{d \mu}{2\pi}\, \left(1-e^{-\beta\hbar\mu}\right) S_K(\mu)\,e^{-i\mu u}  
\int_{0}^{\infty} dv
\int_{-\infty}^{\infty}\frac{d \nu}{2\pi} \left(1-e^{-\beta\hbar\nu}\right) S_K(\nu)\,e^{-i\nu v}
\\ \nonumber
&& \quad\times  \alpha(s-u+v)\, e^{J(s-u+v)}\,
\sum_{k=-\infty}^{\infty}J_k(a)^2 
\left( e^{\frac{i}{\hbar}eV_{dc}(s-u+v)}\, e^{ik\Omega(s-u+v)}   +\hbox{c.c.}\right) .
\end{eqnarray}

We now insert the  Fourier representation (\ref{FFalphaJ}) of $\alpha(t)e^{J(t)}$ into appropriate $s$-dependent terms of Eq.~(\ref{C0ofs2}) which then takes the form
\begin{eqnarray}\label{C0ofs3} 
&&C_0(s)  =   \frac{\hbar^2C^2}{e^2}\int_{-\infty}^{\infty}\frac{d\omega}{2\pi}\,\omega^2\ S_K(\omega)\, e^{-i\omega s}  \\  \nonumber
&& \quad
+\frac{C^2}{e^2}\int_{0}^{\infty} du \, \int_{-\infty}^{\infty}\, \frac{d \mu}{2\pi}\, \left(1-e^{-\beta\hbar\mu}\right) S_K(\mu)\,e^{-i\mu u}\int_{0}^{\infty} dv\, \left(\alpha(v) 
 e^{J(v)} -\hbox{c.c.}\right) \\ \nonumber
&& \qquad\times
\int_{-\infty}^{\infty}\, \frac{d \nu}{2\pi}\, S_K(\nu)\, e^{-i\nu( s+u)}\left[1-e^{-i\nu v}\right]  \sum_{k=-\infty}^{\infty}J_k(a)^2
\left( e^{\frac{i}{\hbar}eV_{dc}v}e^{ik\Omega v} +\hbox{c.c.}\right)
\\ \nonumber
&& \quad +\frac{C^2}{e^2}\int_{0}^{\infty} du \, \int_{-\infty}^{\infty}\, \frac{d \mu}{2\pi}\, \left(1-e^{-\beta\hbar\mu}\right) S_K(\mu)\,e^{-i\mu u}\int_{0}^{\infty} dv\, \left(\alpha(v)  e^{J(v)}-\hbox{c.c.}\right) \\ \nonumber
&& \qquad\times
\int_{-\infty}^{\infty}\, \frac{d \nu}{2\pi}\, S_K(\nu)\, e^{-i\nu( s-u)}\left[1-e^{i\nu v}\right]  \sum_{k=-\infty}^{\infty}J_k(a)^2
\left( e^{\frac{i}{\hbar}eV_{dc}v}\, e^{ik\Omega v}+\hbox{c.c.}\right)
\\ \nonumber
&&\quad
-\frac{ C^2}{e^2} \int_{0}^{\infty} du\,  
\int_{-\infty}^{\infty}\, \frac{d \mu}{2\pi}\, \left(1-e^{-\beta\hbar\mu}\right) S_K(\mu)\,e^{-i\mu u}  
\int_{0}^{\infty} dv\, 
\int_{-\infty}^{\infty}\, \frac{d \nu}{2\pi}\, \left(1-e^{-\beta\hbar\nu}\right) S_K(\nu)\,e^{-i\nu v}
\\ \nonumber
&& \qquad \times  \int_{-\infty}^{\infty}  \frac{d\omega}{2\pi}\, F(\omega) e^{-i\omega (s-u+v)} 
\sum_{k=-\infty}^{\infty}J_k(a)^2
\left( e^{\frac{i}{\hbar}eV_{dc}(s-u+v)}\, e^{ik\Omega(s-u+v)}   +\hbox{c.c.}\right) .
\end{eqnarray}
From this result we can straightforwardly extract the spectral function $S(\omega)$ as
\begin{eqnarray}\label{Somega}
&&S(\omega)  =   \frac{\hbar^2C^2}{e^2}\,\omega^2\ S_K(\omega)  \\  \nonumber
&& \quad
+\frac{C^2}{e^2}\int_{0}^{\infty} du \, \int_{-\infty}^{\infty}\, \frac{d \mu}{2\pi}\, \left(1-e^{-\beta\hbar\mu}\right) S_K(\mu)\,e^{-i\mu u}\int_{0}^{\infty} dv\, \left(\alpha(v) 
 e^{J(v)} -\hbox{c.c.}\right) \\  \nonumber
&& \qquad\times
 S_K(\omega)\, e^{-i\omega u}\left[1-e^{-i\omega v}\right]  \sum_{k=-\infty}^{\infty}J_k(a)^2
\left( e^{\frac{i}{\hbar}eV_{dc}v}e^{ik\Omega v} +\hbox{c.c.}\right)
\\ \nonumber
&& \quad +\frac{C^2}{e^2}\int_{0}^{\infty} du \, \int_{-\infty}^{\infty}\, \frac{d \mu}{2\pi}\, \left(1-e^{-\beta\hbar\mu}\right) S_K(\mu)\,e^{-i\mu u}\int_{0}^{\infty} dv\, \left(\alpha(v)  e^{J(v)}-\hbox{c.c.}\right) \\ \nonumber
&& \qquad\times
 S_K(\omega)\, e^{i\omega  u}\left[1-e^{i\omega v}\right]  \sum_{k=-\infty}^{\infty}J_k(a)^2
\left( e^{\frac{i}{\hbar}eV_{dc}v}\, e^{ik\Omega v}+\hbox{c.c.}\right)
\\ \nonumber
&&\quad
-\frac{ C^2}{e^2} \int_{0}^{\infty} du\,  
\int_{-\infty}^{\infty}\, \frac{d \mu}{2\pi}\, \left(1-e^{-\beta\hbar\mu}\right) S_K(\mu)\,e^{-i\mu u}  
\int_{0}^{\infty} dv\, 
\int_{-\infty}^{\infty}\, \frac{d \nu}{2\pi}\, \left(1-e^{-\beta\hbar\nu}\right) S_K(\nu)\,e^{-i\nu v}
\\ \nonumber
&& \qquad \times e^{i\omega(u-v)}\sum_{k=-\infty}^{\infty}J_k(a)^2\,
\left[ F(\omega+eV_{dc}/\hbar+k\Omega)+  F(\omega-eV_{dc}/\hbar-k\Omega)\right] .
\end{eqnarray}
\end{widetext}
Using the symmetry (\ref{symSK}) of $S_K(\omega)$ and the symmetries
$
\alpha^*(t)=\alpha(-t)$, $ J^*(t)=J(-t)
$
following from Eqs.~(\ref{Joft}) and (\ref{ThetaTheta4}), one can show that the second and third terms of Eq.~(\ref{Somega}) are complex conjugate and that the spectral function $S(\omega)$ is real.
Accordingly, the result (\ref{Somega}) may be written as the sum of three terms (\ref{Somegasum}) given in the main text with noise components $S_N(\omega)$, $S_T(\omega)$ and $S_{NT}(\omega)$ specified in Eqs.~(\ref{SNomega}) -- (\ref{SNTomega}).

%%%%%%%%%%%%%%%%%%%%%%%

\begin{acknowledgments}
The authors wish to thank Carles Altimiras, Daniel Est\`eve, Philippe Joyez, Thierry Martin, Christophe Mora and Fabien Portier for helpful discussions and correspondence.
\end{acknowledgments}

%%%%%%%%%%%%%%%%%%%%
%%%%%%%%%%%%%%%%%%%%%%
%%%%%%%%%%%%%%%%%%%%%
%%%%%%%%%%%%%%%%%%%%%%
%%%%%%%%%%%%%%%%%%%%%

\end{document}